\documentclass{aa}  
\usepackage{graphicx}
\usepackage{txfonts}
\usepackage{hyperref}
\usepackage{amsmath,amstext}
\usepackage[T1]{fontenc}
\usepackage{ae,aecompl}
\usepackage{graphicx}
\usepackage{amssymb}
\usepackage{booktabs,caption}
\usepackage{subcaption}
\usepackage[normalem]{ulem}
\usepackage{float}
\usepackage[dvipsnames]{xcolor}

\newcommand{\ODG}[1]{\,$\times 10^{\,#1}$}

\newcommand{\Xhii}{x_{\text{H\fontsize{5}{6}\selectfont II}}}
\newcommand{\indexXhii}{x_{\text{H\fontsize{5}{6}\selectfont II},i}}
\newcommand{\Msol}{M$_{\odot}$}

\begin{document} 
\newcommand{\AD}[1]{\textcolor{red}{(Aristide: #1)}}
\newcommand{\red}[1]{#1}
\newcommand{\BS}[1]{\textcolor{green}{(Benoit: #1)}}
\newcommand{\err}{\operatorname{MSE}}
\newcommand{\errBSD}{\err\left(\log_{10}\left[V^{2}\frac{dn_{b}}{dV}\right]\right)}
\newcommand{\errMofV}{\err\left(\log_{10}\left[M_{coll}(V_{\text{ion}})\right]\right)}

   \title{A bubble size distribution model for the Epoch of Reionization}


   \author{Aristide Doussot
          \inst{1}\fnmsep\inst{2}
          \and
          Benoît Semelin\inst{1}
          }

   \institute{Sorbonne Universit\'e, Observatoire de Paris, PSL research university, CNRS, LERMA, F-75014 Paris, France
                \and
              Sorbonne Université, UMR7095, Institut d’Astrophysique de Paris, 98 bis Boulevard Arago, F-75014, Paris, France
             }


 
  \abstract
   {}
    {The bubble size distribution is a summary statistics that can be computed from the observed 21-cm signal from the Epoch of Reionization. As it depends only on the ionization field and is not limited to gaussian information, it is an interesting probe, complementary to the power spectrum of the full 21-cm signal. Devising a flexible and reliable theoretical model for the bubble size distribution paves the way for using it for astrophysical parameters inference.}
    {The proposed model is built from the excursion set theory and a functional relation between the bubble volume and the collapsed mass in the bubble. Unlike previous models it accommodates any functional relation or distributions. Using parameterized relations allows us to test the predictive power of the model by performing a minimization best-fit to the bubble size distribution obtained from a high resolution, fully coupled radiative hydrodynamics simulations, HIRRAH-21.    }
   { Our model is able to provide a better fit to the numerical bubble size distribution at ionization fraction of $\Xhii\sim 1\%$ and $3\%$ than other existing models. Moreover, the bubble volume to collapsed mass relation corresponding to the best-fit parameters, which is not an observable, is compared to numerical simulation data. A good match is obtained, confirming the possibility to infer this relation from an observed bubble size distribution using our model. Finally we present a simple algorithm that empirically implements the process of percolation. We show that it extends the usability of our bubble size distribution model up to $\Xhii \sim 30\%$.}
   {}


   \maketitle
%
\section{Introduction}

The Epoch of Reionization (EoR) is the era when the formation of the first stars and galaxies triggered a major transformation of the intergalactic medium (IGM): from the cold and neutral state that followed recombination (at $z\sim 1000$), it transitions to a warm ionized state. This transition is brought about by the emission of ionizing radiations from the first sources. The resulting ionized bubbles first grow and then percolate until the last islands of neutral hydrogen finally disappear. While the first star in the observable universe may have formed as early as $z \sim 65$ \citep{Naoz2006}, the beginning of the reionization process, defined as the time when it becomes observable (through non-negligible contribution to the Thomson scattering optical depth, the $21$-cm signal or individual high redshift proto-galaxies) is likely in the $z=15-30$ range. A more accurate timing depends on the very uncertain efficiency of the formation of the first stars that rely on a number of physical processes such as molecular Hydrogen formation and dissociation, the dynamical feedback from the first supernovae explosions and the subsequent chemical enrichment of the gas. The end of the process, that is the complete ionization of the IGM, occurs around $z \sim 5.5 - 6.5$ \citep[e.g.][]{Fan2006, McGreer2015, Gangolli2021,Qin2021,Bosman2022}. Among the many possible probes of the EoR (galaxy luminosity functions, quasar proximity effects, Gamma-ray bursts, Lyman-alpha emitters observations) the 21-cm signal emitted by the neutral IGM holds a unique place. Indeed, rather than sampling the universe on individual lines of sight, it can provide a full 3D tomographic mapping of the EoR \citep[see e.g.][for reviews]{Furlanetto2006, Mellema2013}. While 21-cm signal statistics such as the global signal or the power spectrum are in principle observable with current instruments if systematics are brought under control \citep[e.g.][]{Bernardi2009,Bernardi2010,Ghosh2011,Yatawatta2013,Jelic2014,Asad2015,Remazeilles2015,Offringa2016,Procopio2017,Line2017}, only the Square Kilometer Array should have sufficient sensitivity to perform tomography at $5 \arcmin$ angular resolution \citep{Koopmans2014}.

The difficulty with interpreting observations of the 21-cm line is that it depends on several local quantities: the gas density, ionization fraction, kinetic temperature and velocity, and  the local flux in the Lyman-alpha line for the coupling of spin temperature to the kinetic temperature. These values are in turn determined by non-local and non-linear processes: mainly gravity and radiative transfer, but also all the processes that regulate the formation of the sources of radiations. Consequently, one must build parameterized models, theoretical, semi-numerical or numerical, that attempt to account for all these effects and explore the parameter space in the hope of matching the observed signal. Obviously, the methodology to explore the parameter space efficiently (beyond brute force griding) has been an active field of research. For example, \citet{Pober2014} use the Fisher matrix formalism in combination with the $21$cmFAST semi-numerical code \citep{Mesinger2011} to estimate the constraints that can be obtained from future observations with the HERA interferometer. \citet{Greig2015a,Greig2017a, Greig2018a} deploy a full Bayesian Markov-Chain-Monte-Carlo (MCMC) framework also based on 21cmFAST to compute the full posterior distribution for the model parameters. As even using a semi-numerical model at low resolution in conjunction with a MCMC approach represents a significant computational cost, some authors have explored using emulators of the semi-numerical model, using Gaussian Processes \citep{Kern2017} or Neural Networks \citep{Schmit2018,Jennings2018,Cohen2020,Bye2021,Bevins2021}. Another avenue of research consists in training supervised learning algorithms to perform the inverse modelling that retrieve the model parameters from the signal \citep{Shimabukuro2017,Gillet2018,Doussot2019,Hassan2020,Zhao2021,Zhao2022,Prelogovic2022}. While these works were limited to maximum-likelihood type inference, the posterior distribution could in principle be accessed using Bayesian Neural Network \citep[see, e.g.][]{Hortua2020} or with density estimators also using neural network for dimensionality reduction \citep{Zhao2021}.

Most of these parameter inference studies use the 21-cm signal power spectrum as a metric of the difference between the model and reality, either as a proof of concept, as in the above references, or based on the existing upper limits \citep{Mondal2020,Ghara2020,Ghara2021,Greig2021,Greig2021a,Abdurashidova2022}. Others have used the global signal \citep[e.g.][]{Monsalve2019} or, as a proof of concept only, the full tomography \citep[e.g.][]{Gillet2018,List2020,Zhao2021}. When using the global signal or the power spectrum, fluctuations induced by the ionisation fraction, the X-ray heating and the Lyman-$\alpha$ coupling are mixed. Only at times when one type dominates over the others can we hope to isolate it. Tomography, on the other hand, offers the possibility to disentangle the different contributions, for example by identifying individual ionized bubbles. However, as it is not a summary statistics, it suffers from a high level of thermal noise. This is why a summary statistics derived from the tomography and associated to a single source of fluctuation such as the Bubble Size Distribution (BSD), that depends only on the ionization field, may be an interesting metric for parameter inference. It would likely provide constraints on the parameter space with different degeneracies than the 21-cm power spectrum for example. It is also sensitive to the non-Gaussian information in the signal. While instrumental limitations, starting with the increasing thermal noise at higher angular resolution \citep[e.g.][]{Mellema2013}, will affect the process, methods to obtain the bubble size distribution from tomographic observations have been explored \citep{Giri2018,Bianco2021}. Semi-numerical, or even numerical simulations could then be used to perform parameter inference based on this quantity. If an accurate theoretical model can be devised however, it stands a good chance of involving a lower computational cost during the inference.

To our knowledge, there exists currently only one theoretical model: it is described in \citet{Furlanetto2004a} (hereafter FZH04) and relies on the excursion set theory (see also \citet{Paranjape2014,Paranjape2016} for some improvements to the model). The FZH04 model has been tested against numerical simulations in \citet{Mesinger2007} and \citet{Lin2016} who find some level of discrepancy. The fundamental assumption in the FZH04 model is that the mass of the ionized gas in each bubble is uniquely determined and directly proportional to the collapsed halo mass in the bubble. This is actually a necessary assumption for the model to be able to deliver an analytic formula. While this assumption is natural, it is also simplistic. Indeed, environmental effects such as source clustering, inhomogeneous recombination or external photo-evaporation, are bound to introduce some scattering into the one-to-one relation, but also induce a non-linear relation. Another limitation of the model is associated to the percolation process. Indeed, the excursion set formalism evaluates the probability for a region to be ionized, but does not account for the possibility that such regions overlap. \citet{Furlanetto2016} have studied the process of percolation for the ionization field and shown that an ionized region with infinite volume (that is extending from one side of the box to the other in simulations) appears as soon as the average ionized fraction reaches $\Xhii\sim 0.1$. There is no framework in the FHZ04 model to account for percolation.

In this work, we present an alternative, more flexible theoretical model to describe the BSD. By reverting to the original Press-Schechter spirit rather than using the excursion set theory, we lose something in terms of formal rigor, but we gain much in term of flexibility. Not only are we able to consider any functional relation between the ionized gas mass and collapsed halo mass, but we can also move from a one-to-one relation to a distribution characterized by its average and dispersion. We will show that this more flexible representation offers a good match to the result of a high-resolution numerical simulation. Then, a choice of an astrophysically-motivated parameterization for the relation opens the way for parameter inference using the BSD as a metric. 

In section 2 we will present our methods: the reference numerical simulation, the algorithm for computing the numerical BSD and our proposed theoretical model. In section 3 we test the predictive power of the model by performing a best fit of the theoretical model parameters to the numerical BSD. The resulting ionized-gas-volume to collapsed-mass relation corresponding to the best-fit parameters is then compared to the same relation directly computed from the simulation data. Section 4 evaluates the improvements produced by including a dispersion in the relation. Section 5 presents and evaluates the performances of an empirical percolation algorithm. Finally section 6 gives our conclusions.

\section{The bubble size distribution: Numerical computation and analytical models }
\subsection{The HIRRAH-21 simulation}\label{sSec:HIRRAH21}

    The High Resolution Radiative Hydrodynamics for 21-cm (HIRRAH-21) simulation was performed using the \textsc{licorice} code, detailed in \citet{Baek2009} (see \citet{Baek2010, Semelin2016} for subsequent upgrades). \textsc{licorice} computes the evolution of gas, dark matter and radiation on cosmological scales. It is a particle-based code modeling the dynamics using the Tree-SPH method  \citep[and references therein]{Semelin2002}. The dynamics is fully-coupled to the radiative transfer of ionizing UV and X-rays using a Monte-Carlo scheme described in \citet{Baek2009, Baek2010}. Notable features easily implemented in the Monte Carlo approach are that the photons are propagated with the correct speed of light on an adaptive grid, and that cosmological redshifting is included. This is especially relevant for handling X-rays that can travel long distances before absorption.
    
    HIRRAH-21 is meant to complement the 21SSD set of simulations presented in \citet{Semelin2017a} and whose 21-cm signal predictions are available at \href{https://21ssd.obspm.fr/}{https://21ssd.obspm.fr}. The initial conditions for HIRRAH-21 were produced with the \textsc{music} package \citep{Hahn2011} using the same random seeds as for the 21SSD simulations for the shared resolution levels, and a new one for the additional resolution level in HIRRAH-21 (see below). Thus the initial density field in HIRRAH-21 differs from those in 21SSD only by fluctuations at the newly resolved scales. For reference, the X-ray production parameters are $f_X=1$ and $r_{H/S}=0.5$ (see \citet{Semelin2017a} for definitions), although in this work, where we focus on the ionized bubbles size distribution, such moderate values have little impact on the results. 
    
    HIRRAH-21 follows the evolution of 2048$^{3}$ particles, half of them baryons and the other half dark matter, in a 200 $h^{-1}$cMpc box. A parallelized halo finder based on a friend-of-friend scheme resolves halos down to $4$\ODG{9} \Msol  (that is $20$ dark matter particles). The dynamical timestep is $0.5$ Myr (divided by 3 at expansion factor smaller than $0.03$) and the gravitational softening is $\epsilon = 2$ ckpc. The parameters of star formation and the ionizing photon escape fraction are the same as in 21SSD, yielding an earlier reionization history due to the better mass resolution for halos. The spectral properties of the sources are also the same as in 21SSD, resulting from a Salpeter initial mass function truncated at $1.6$ M$_{\odot}$ and $120$ M$_{\odot}$. Roughly $5 \times 10^{11}$ Monte Carlo photon packets were propagated during the simulation, with $7 \times 10^{10}$ in-flight photons by the end of reionization which occurs around $z\sim 6.9$.
    
    \begin{figure}[t]
    \begin{center}
    \includegraphics[width=\columnwidth]{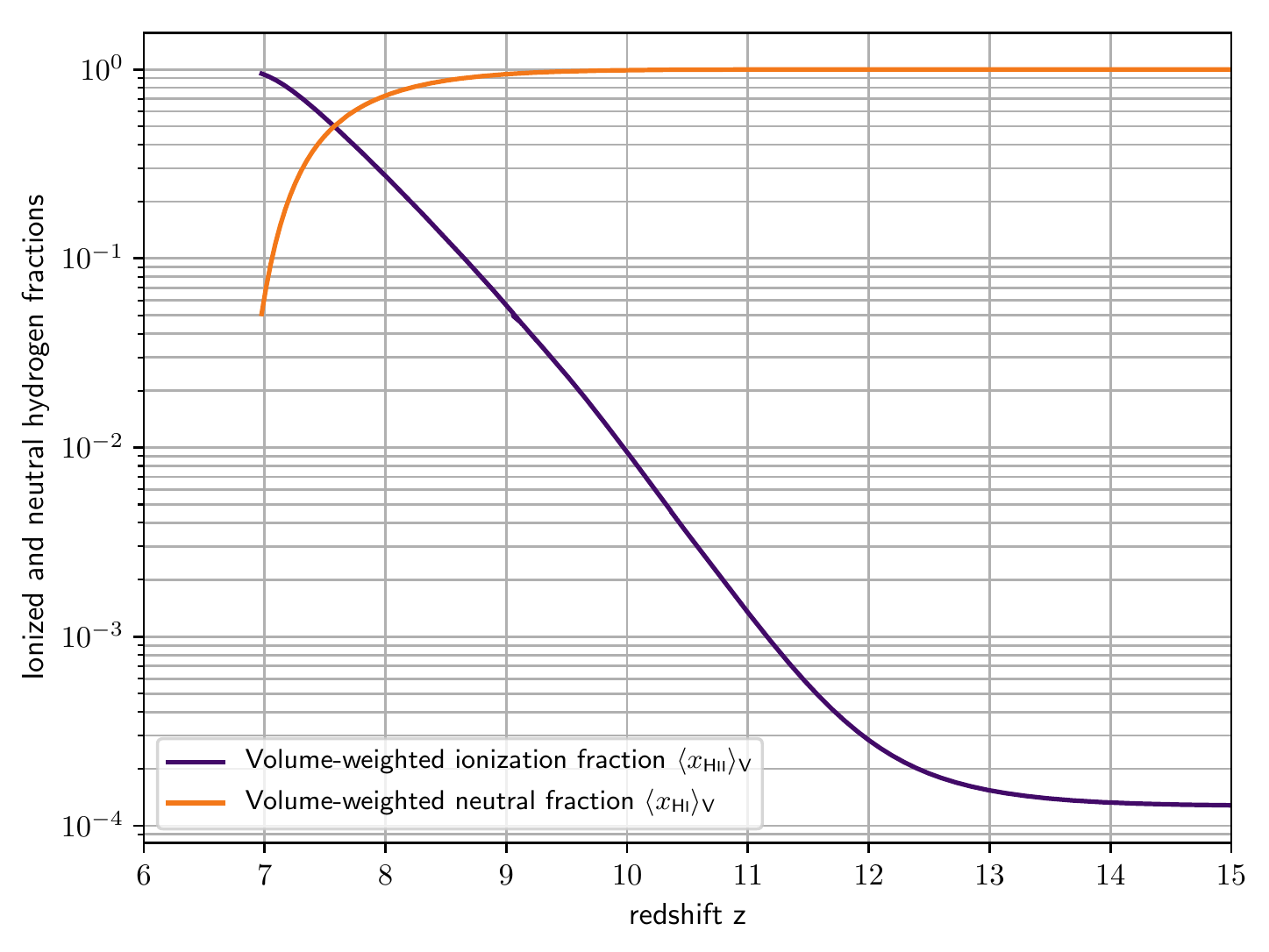}
    \caption{Volume-weighted ionized (purple) and neutral (orange) fraction as a function of the redshift from the HIRRAH-21 simulation.}
    \label{Fig:XionHIRRAH}
    \end{center}
    \end{figure}
    
    For reference, we show in Fig.\ref{Fig:XionHIRRAH} the volume-weighted ionized (purple) and neutral (orange) fractions as functions of the redshift from the HIRRAH-21 simulation. The optical depth obtained from this simulation is $\tau \sim 0.0692$. We use a standard $\Lambda$-CDM cosmology with parameter values in accordance with \citet{Planck2016}: $H_{0}$=$67.8$km.s$^{-1}$, $\Omega_{\text{M}}$=$0.308$, $\Omega_{\text{b}}$=$0.0484$,  $\Omega_{\Lambda}$=$0.692$, $\sigma_{8}$=$0.8149$ and $n_{\text{s}}$=$0.968$. HIRRAH-21 was performed on $16386$ CPU cores using 4096 MPI domains and requiring $\sim 3$ \ODG{6} CPU hours.

\subsection{Numerical computation of the bubble size distribution}

    To identify ionized bubbles in HIRRAH-21, we use a friend-of-friend method, based on the ionization field interpolated on a 1024$^{3}$ grid. It means that we can detect bubbles with a characteristic size of $\gtrsim 288$ ckpc. We tag as ionized any cell that has an ionization fraction above a given threshold \red{$\Xhii^{\text{threshold}}$}. A bubble is a connected set of ionized cells. A similar algorithm has been used and analyzed in previous studies \citep{Friedrich2011, Furlanetto2016, Giri2018}. 
    
    When the bubble is identified, we compute its volume as :$V_{\text{bubble}}=V_{\text{cell}}\sum_{i=0}^{N_{\text{cell}}} \indexXhii$ where $V_{\text{cell}}$ is the volume of one cell, $N_{\text{cell}}$ the number of cells in the considered ionized bubble and $\indexXhii$ the ionization fraction of cell $i$. If necessary we can then compute an equivalent radius from the volume assuming a spherical shape. Other methods like the mean-free-path or the spherical-average methods give similar but not strictly equivalent results \citep{Mesinger2007, Zahn2007,Giri2018}.
    
    We subsequently produce a list of bubble volumes which can easily be converted into a number density of bubble in an interval of radius $[r,r+dr]$ : $n_{b}(r)dr$. In accordance with previous works \citep{Furlanetto2016, Giri2018}, to depict the bubble size distribution we will use the quantity $V^{2}\frac{dn_{b}}{dV}$. This quantity measures the fraction of space occupied by ionized bubbles with volume $V$ in a given logarithmic bin of volume $dlnV$. This is equivalent to the probability density function $\frac{dP}{dlnr}=\frac{V}{Q}\frac{dn}{dlnr}$ given in some works \citep{Furlanetto2004a,Lin2016, Giri2019} up to a constant $\frac{3}{Q}$ where Q is the global volume-filling fraction of ionized regions.
    
\begin{figure}[t]
\begin{center}
\includegraphics[width=\columnwidth]{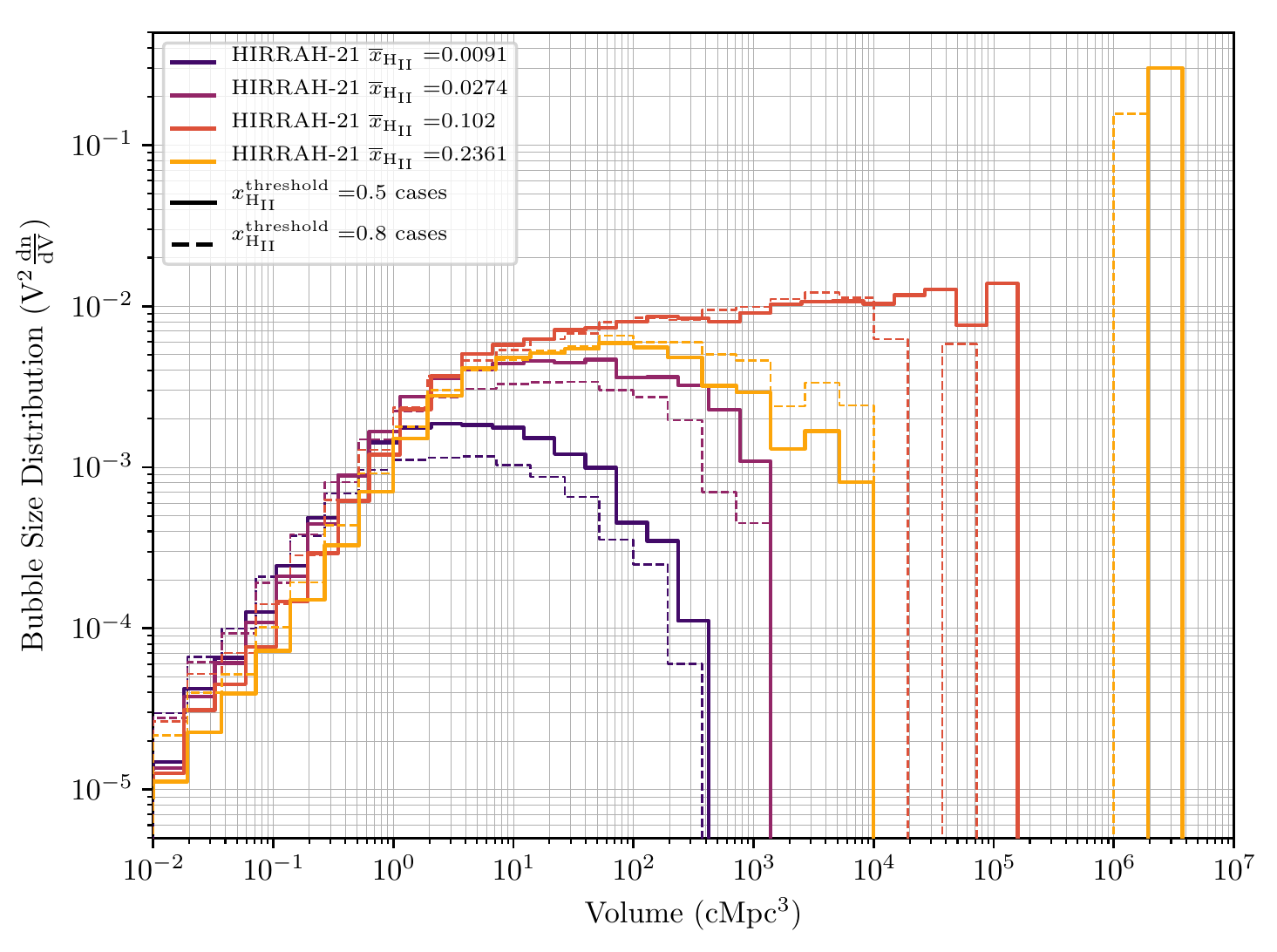}
\caption{Bubble size distribution from the HIRRAH-21 simulation at global ionization fractions $\Xhii\sim$ $1\%$, $3\%$, $10\%$, and $20\%$ \red{for ionization thresholds $\Xhii^{\text{threshold}} = 0.5$ (solid lines) and $0.8$ (dashed lines)}.}
\label{Fig:NumBSD}
\end{center}
\end{figure}
    
        In Fig.\ref{Fig:NumBSD}, we show the bubble size distribution from our HIRRAH-21 simulation at average ionization fractions of $\Xhii\sim$ $1\%$, $3\%$, $10\%$, and $20\%$, which in our simulation correspond respectively to redshifts $z\sim 9.8$, $9.3$, $8.5$, and $8.1$\red{, for ionization thresholds $\Xhii^{\text{threshold}} = 0.5$ (solid lines) and $0.8$ (dashed lines)}. At large volumes (V$\gtrsim 10$cMpc$^{3}$), the distribution drops for $\Xhii\sim$ $1\%$ and $3\%$ because we consider very early stages of reionization when bubbles have not yet had time to grow. The distribution is more or less flat for $\Xhii\sim$ $10\%$, and drops again for $\Xhii\sim$ $20\%$. \red{The decrease in amplitude after $\Xhii\sim$ $10\%$ is theoretically expected because of the percolation process that merges large bubbles into a single one spanning of whole simulation box \citep{Furlanetto2016}. The average ionization value at which this percolated cluster forms seems to be only weakly model dependant as shown in \citet{Furlanetto2016}. In our case it appears at $\Xhii\sim$ $10\%$}. For smaller bubbles (V$\lesssim 5$cMpc$^{3}$), the finite resolution of our simulation affects the results leading to a depletion in the number of detected bubbles.
        
        \red{Fig.\ref{Fig:NumBSD} also shows that the BSD obtained for $\Xhii^{\text{threshold}}=0.8$ is similar to the one using a threshold of $0.5$, with a slightly delayed evolution. This is understandable as the threshold value simply modifies our definition of what is part of the ionized bubbles but does not affect the physical processes that rule their evolution. The exact same bubble will have, at the same redshift, an estimated volume smaller in the $\Xhii^{\text{threshold}}=0.8$ case than in the $\Xhii^{\text{threshold}}=0.5$ case or, equivalently, a volume that was reached somewhat earlier in the $\Xhii^{\text{threshold}}=0.5$ case. Note however that the changes in the BSDs created by changing the threshold value are smaller than those created by the evolution between two of our sampled average ionization fractions and thus would correspond to very small shifts in the timing of the ionization history. Moreover, the parameters that will be inferred from the BSD in section 3 do not seem to evolve much. Thus, in the following study, we will only use $\Xhii^{\text{threshold}}=0.5$. }
        
\subsection{The excursion set based model of the bubble size distribution}

    An analytical model to compute the bubble size distribution has been proposed by \citet{Furlanetto2004a} (already referred to as FHZ04), based on the assumption that the mass of the gas in an ionized region is directly proportional to the mass of the collapsed objects inside the region, the proportionality factor being called the efficiency factor $\zeta$. We refer the reader to FHZ04 for the full details on the model. The end result is that the bubble size distribution is expressed analytically as :
    \begin{equation}
        m\frac{dn}{dm}=\sqrt{\frac{2}{\pi}}\frac{\overline{\rho}}{m}\left\lvert\frac{d\ln \sigma}{d\ln m}\right\rvert\frac{B_{0}}{\sigma (m)}\exp\left[-\frac{B^{2}(m,z)}{2\sigma^{2}(m)}\right]
    \label{Eq:BSD_Furlanetto}
    \end{equation}
    \noindent
    where $m$ is the ionized gas mass in the bubble, $dn$ is the number density of bubbles with gas mass between $m$ and $m+dm$, $\overline{\rho}$ is the mean density of the universe, $\sigma (m)$ is the mass variance of the linear density field and $B(m,z)=B_{0}+B_{1}\sigma^{2}(m)$ is a linear fit to the overdensity threshold needed so that the collapsed fraction in the region is sufficient to ionize it. To enable a comparison with our own model, we will convert the mass-dependency of the distribution in Equation \ref{Eq:BSD_Furlanetto} into a size dependency using the relation : $$m=V\,\overline{\rho}\left( 1+B(m,z)\right).$$
    Using a formalism derived from the extended Press-Schechter theory, this model relies on the assumption that the mass of ionized gas depends linearly on the mass of collapsed objects to produce an analytical formulation for the BSD. Thus it cannot take into account another functional relation nor a dispersion in the relation.

\subsection{A new, flexible model for the bubble size distribution}

\subsubsection{Computing the conditional mass function}\label{sSec:Theory_CMF}

	Our model for computing the BSD aims at using a flexible physical prescription to relate, for a given {\sl ionized} region, the amount of mass bound in self-gravitating objects, i.e the collapsed mass, to the volume of the region. So the first task is to describe the population of collapsed objects in a given region of the universe. Since ionized bubbles typically form in overdense regions, one cannot simply use the Halo Mass Function (HMF) formalism, but we must resort to the conditional mass function (CMF) first presented in \citet{Lacey1993,Lacey1994}.
	
	Let us consider a region of volume $V_{0}$, mass $M_{0}=V_{0}\overline{\rho}$ and mass variance $\sigma_{0}=\sigma(M_{0})$. In the following study, $\sigma (M)$ is computed using \textsc{TheHaloMod} online calculator \citep{Murray2013,Murray2020}.By construction, our model can use any form for the conditional mass function. In this study we will mainly use the conditional mass function derived from the empirical best-fitting halo mass function of \citet{Sheth1999}. For a region with overdensity $\delta_{0}$, the CMF can be written \citep{Rubino-Martin2008}:
    \begin{equation}
        n_{c,\text{ST}}=\sqrt{\frac{2}{\pi}}\frac{\overline{\rho}}{m}\left\lvert\frac{d\sigma}{dm}\right\rvert \frac{\sigma (m)\left\lvert T(\sigma,\sigma_{0}) \right\rvert }{(\sigma(m)-\sigma_{0})^{\frac{3}{2}}}\exp\left[-\frac{(B_{\text{ec}}(\sigma,z)-\delta_{0})^{2}}{2(\sigma(m)-\sigma_{0})^{2}}\right]
    \label{Eq:CMF_ST}
    \end{equation}
    where $$T(\sigma,\sigma_{0})=\sum_{n=0}^{5}\frac{(\sigma_{0}^{2}-\sigma(m)^{2})^{n}}{n!}\frac{\partial^{n}(B_{\text{ec}}(\sigma,z)-\delta_{0})}{\partial(\sigma^{2})^{n}}$$ with $B_{\text{ec}}$ the barrier derived by \citet{Sheth2001} and given by $$B_{\text{ec}}(\sigma,z)=\sqrt{0.707}\delta_{c}(z)\left[ 1+0.485\left( \frac{0.707\delta_{c}(z)^{2}}{\sigma^{2}} \right)^{-0.615}\right].$$
	
    In Section \ref{sSec:CMF_sensitivity} we will compare the results with those for the CMF derived in the extended Press-Schechter formalism \citep{Press1974,Bond1991,Lacey1993,Lacey1994} to highlight our model sensitivity to this initial choice for the CMF. This other form of the conditional mass function can be written as :
    \begin{equation}
        n_{c,\text{EPS}}=\sqrt{\frac{2}{\pi}}\frac{\overline{\rho}}{m}\left\lvert\frac{d\sigma}{dm}\right\rvert\frac{\sigma (m)(\delta_{c}-\delta_{0})}{(\sigma(m)-\sigma_{0})^{\frac{3}{2}}}\exp\left[-\frac{(\delta_{c}-\delta_{0})^{2}}{2(\sigma(m)-\sigma_{0})^{2}}\right]
    \label{Eq:CMF_EPS}
    \end{equation}
    where $\delta_{c}=1.686$. 
    
\subsubsection{The collapsed mass probability distribution }\label{sSec:PMcoll}

	 As long as the non-gaussianities resulting from non-linear structure formation have not had time to develop on the scale of the considered region, the probability density function for a region of volume $V_{0}$ to have an overdensity $\delta_{0}$ is given by:
    \begin{equation}
         P_{V_{0}}(\delta_{0})=\frac{1}{\sqrt{2\pi}\sigma_{0}}\exp\left(-\frac{\delta_{0}^{2}}{2\sigma_{0}^{2}}\right).
    \label{Eq:Pdelta}
    \end{equation}

	Then, the collapsed mass probability density for a region of volume $V_{0}$ can be expressed as a marginalization over $\delta$:
    \begin{equation}
       P_{V_{0}}(M_{coll})=\int_{-\infty}^{\delta_{l}}P_{V_{0}}(M_{coll}\,|\,\delta)P_{V_{0}}(\delta)d\delta
    \label{Eq:PMcoll_marginalization}
    \end{equation}
where $P_{V_{0}}(M_{coll}\,|\,\delta)$ is the conditional probability for a region of volume $V_{0}$ to have a collapsed mass $M_{coll}$ knowing that it has an overdensity $\delta$. In the CMF formalism this conditional probability is a Dirac distribution: the collapsed mass is uniquely determined by the volume and overdensity of the region. This collapsed mass is however only an average value for all regions with identical volume and overdensity. Sample variance of the density field within each region will generate a distribution of values around this average. Thus, we explore two ways to model $P_{V_{0}}(M_{coll}\,|\,\delta)$:
\begin{itemize}
\item The simplest approach is to use the average value from the CMF formalism. In this case we can directly compute the collapsed mass using a generic conditional mass function $n_{c}$ :
    \begin{equation}
        M_{coll}(\delta_{*},V_{0})=V_{0}\int_{M_{min}}^{M_{0}}n_{c}(m,\delta_{*})mdm.
    \label{Eq:Mcoll}
    \end{equation}
    
    The minimum mass threshold $M_{min}$ is necessary in our case since we are interested only in halos that can efficiently form stars and thus produce ionizing photons. It can be set for example at the atomic cooling limit $ \sim 1$\ODG{8} M$_\odot$, or at the resolution limit of a simulation, allowing our model to fit results in various frameworks. For this study we set $M_{min}\sim 4$\ODG{9} \Msol \, corresponding to the resolution limit of the HIRRAH-21 simulation.
    With this assumptions, the conditional probability $P_{V_{0}}(M_{coll}\,|\,\delta)$ is:

$$P_{V_{0}}(M_{coll}'|\delta)=\delta_D\left( M_{coll}'-M_{coll}(\delta_{*},V_{0})\right)$$

where $\delta_D$ is the Dirac distribution and we can rewrite the collapsed mass distribution inside regions of volume $V_{0}$ of Equation \ref{Eq:PMcoll_marginalization} :
    \begin{equation}
       P_{V_{0}}(M_{coll})=P_{V_{0}}(\delta_{*})\frac{d\delta}{dM_{coll}}(M_{coll}(\delta_{*})).
    \label{Eq:PMcoll}
    \end{equation}

\item The more realistic approach is to consider that regions with the same size and average overdensity can have a distribution of collapsed masses. A simple way to take the sample variance within each region into account is to model the number of objects in each mass bin using a Poisson distribution \citep{Barkana2004}. Note that since we are not interested simply in the {\sl total number} of collapsed objects but in the {\sl total mass} in collapsed objects, we cannot work with a single mass bin, relying on the fact that the sum of two independent Poisson distributed random variables is Poisson distributed: indeed this property does not work for a linear combination of independent random variables (the total collapsed mass in our case). 

The Poisson distribution of the number of halos in a region of volume $V_{0}$ and overdensity $\delta_{*}$, with a mass in the range $[M_{*},M_{*}+dM]$, must have an expected value equal to the average number of halos  in this range as predicted by the conditional mass function: $V_{0}n_{c}(M_{*},\delta_{*})dM$. This is sufficient to fully define the Poisson distribution.
Therefore we can numerically construct the distribution $P_{V_{0}}(M_{coll}\,|\,\delta_{*})$ with a Monte Carlo sampling approach. In each mass bin, we draw the number of halos from the correct Poisson distribution, then we combine these values to compute the total collapsed mass, and iterate the process many times to build the total collapsed mass distribution. Since these distributions depend on two parameters ($V_{0}$ and $\delta_{*}$), it is useful to tabulate them in advance.  We can finally apply Equation \ref{Eq:PMcoll_marginalization} to get the collapsed mass distribution marginalized over the overdensity.

\end{itemize}

\begin{figure}[t]
\begin{center}
\includegraphics[width=\columnwidth]{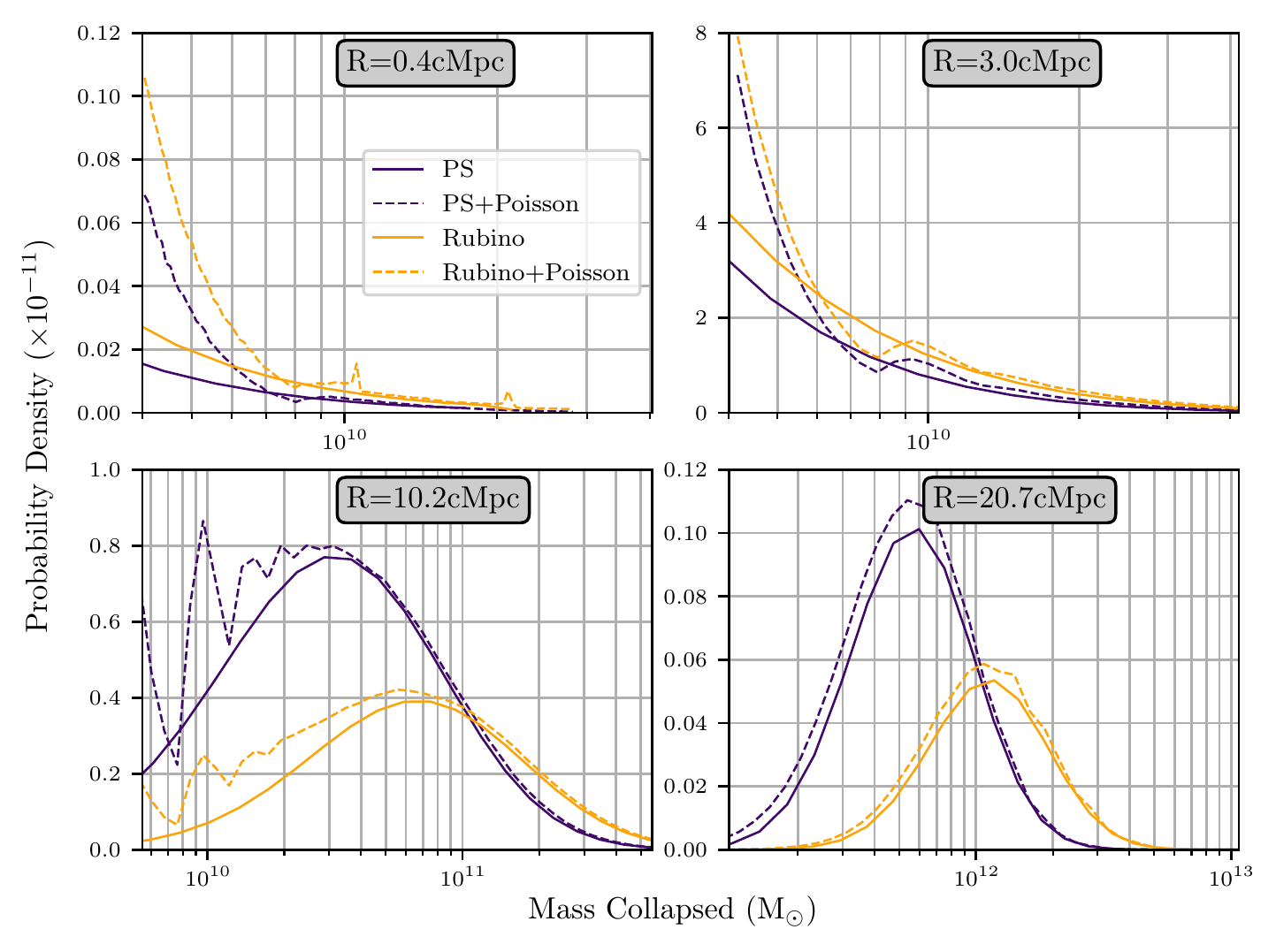}
\caption{Collapsed mass distribution computed at redshift $z\sim9.8$ (corresponding to $\Xhii\sim 1\%$) considering the sample variance (dashed lines) or not (solid lines) using the conditional mass function of Equation \ref{Eq:CMF_EPS} (purple) or of Equation \ref{Eq:CMF_ST} (yellow). The results are shown for ionized regions of radius of : $0.4$ cMpc, $3$ cMpc, $10$ cMpc, and $20$ cMpc.}
\label{Fig:Pmcoll_Comp}
\end{center}
\end{figure}

Figure \ref{Fig:Pmcoll_Comp} represents the collapsed mass distribution computed at redshift $z\sim9.8$ including or not the sample variance, and using a CMF either from Equation \ref{Eq:CMF_EPS} (purple) or Equation \ref{Eq:CMF_ST} (yellow), for various ionized regions sizes. The main difference induced by sample variance is for values of collapsed mass $\lesssim 10^{10}$\Msol, that is values close to our simulation mass resolution of $4$\ODG{9}\Msol. This is expected since in the cases when many bin are populated with halos (thus yielding a large collapsed mass), the central limit theorem applies and the distribution peaks around this average. Furthermore as, at a given size, ionized regions are likely to be regions with a large amount of collapsed mass, only the right-end of the distributions are relevant for the BSD computation. In this range, the only expected difference between BSD computed with or without sample variance will be for the smallest ionized bubbles, where only a few halos a present in the region. The low mass oscillations that can be seen in the cases when the Poisson distribution is used are generated by using a sharp lower mass limit for halos. Indeed, between $M_{min}$ and $2 \times M_{min}$ the region contains at most $1$ halo, with decreasing probability toward larger mass (as dictated by the CMF). When the collapsed-mass reaches $2 \times M_{min}$, suddenly configurations with 2 halos become possible, creating a sharp rise in the probability. Such a phenomenon occurs again, with decreasing amplitude, when integer numbers of $M_{min}$ are reached.

	In Fig.\ref{Fig:Pmcoll_Comp} we also see that the collapsed mass probability distribution peak is shifted to lower collapsed mass when using the CMF based on \citet{Press1974} compared to the \citet{Sheth1999} case, thereby showing the effect on the collapsed mass probability distribution of this well-known difference between the two theories \citep{Sheth1999,Rubino-Martin2008}. This difference impacts the resulting BSD as will be discussed later.\\

\subsubsection{Parameterizable bubble size distribution}

	Having computed the collapsed mass distributions for a wide range of region volumes using Equation \ref{Eq:PMcoll_marginalization}, we can now express the probability for a region of volume $V_{0}$ to be ionized as :
    \begin{equation}
        P_{\text{ion}}(V_{0})=P_{V_{0}}\left(M_{coll}\right)\frac{dM_{coll}}{dV}(V_{0})
    \label{Eq:Pion_NoDispersion}
    \end{equation}
where $\frac{dM_{coll}}{dV}$ is computed from the physical relation $M_{coll}(V_{\text{ion}})$ that links the  volume of an ionized bubble with the collapsed mass it contains. Note that this relation cannot be derived from an HMF formalism since it must implement the condition that the region is ionized. One of the key properties of our model is that it allows to use various physical relations based on theory or simulations results alike. We will investigate different functional relations. The final step to compute the BSD is to write the number per unit volume of ionized bubbles with volume between $V$ and $V+dV$ as:

\begin{equation}
    n_b(V)={1 \over V} P_{\text{ion}}(V)\,dV.
    \label{Eq:our_BSD}
\end{equation}

This last step is very much in the spirit of the original Press-Schecher theory (even if we do not rely on the corresponding Halo Mass Function). What it lacks in rigour it gains in flexibility in the $M_{coll}(V_{\text{ion}})$ relation. The FHZ04 model follows the more rigorous approach of the extended Press-Schechter formalism, but is limited to a linear $M_{coll}(V_{\text{ion}})$  relation, which is the only case where an analytical solution to the diffusion equation is known.
\subsubsection{Accounting for dispersion in the $M_{coll}(V_{\text{ion}})$ relation}\label{sSec:Th_AddDispersion}

	If the physical relation between the ionized bubble volume and the collapsed mass in the bubble displays a non-negligible dispersion, like for our HIRRAH-21 simulation, we can account for it. Given the  collapsed mass distribution for a region of a volume $V_{0}$, we use the distribution characterizing the collapsed mass dispersion at this same volume $P_{V_{0},\text{ion}}(M_{\text{coll}})$ to obtain a modified distribution encapsulating this scatter :
\begin{equation}
	P_{\text{ion}}(V_{0})=\int_{M_{min}}^{\infty}P_{V_{0}}(M_{\text{coll}})P_{V_{0},\text{ion}}(M_{\text{coll}})\,dM_{\text{coll}}.
	\label{Eq:Pion_Dispersion}
\end{equation}

	It is worth noting that, although we will suppose a gaussian behaviour for our following study, the dispersion distribution can theoretically take any form, be it analytical or numerical. Applying Equation \ref{Eq:our_BSD} to our new distribution we can now compute a bubble size distribution corresponding to a physical relation with dispersion $P_{V_{0},\text{ion}}(M_{\text{coll}})$.

\subsubsection{Model summary}

For clarity sake, we hereby sum up the main computation steps of our model. We compute the number per unit volume $n_b(V_{0})$ of ionized regions of volume $V_0$ as follows :

\begin{enumerate}
    \item We first compute $P_{V_{0}}(\delta_{0})$ the probability density function for a region of volume $V_{0}$ to have an overdensity $\delta_{0}$ using Equation \ref{Eq:Pdelta}.
    \item We compute $P_{V_{0}}(M_{coll}\,|\,\delta)$ the conditional probability for a region of volume $V_{0}$ to have a collapsed mass $M_{coll}$ knowing that it has an overdensity $\delta$. This can be done in two ways :
    \begin{itemize}
        \item Assuming that the numerical sample variance effect within ionized regions is negligible it can be computed using a Dirac distribution and the CMF formalism (Eq. \ref{Eq:Mcoll}).
        \item Taking into account the sample variance effect implies to numerically construct $P_{V_{0}}(M_{coll}\,|\,\delta)$ using a Poisson distribution and a Monte Carlo sampling as detailed in Section \ref{sSec:PMcoll}.
    \end{itemize}
    \item We then marginalize over $\delta$ to obtain the collapsed mass probability density $P_{V_{0}}(M_{coll})$ for a region of volume $V_{0}$ using Equation \ref{Eq:PMcoll_marginalization}
    \item We define a parameterized physical relation $M_{coll}(V_{\text{ion}})$ that connects the  volume of a ionization bubble with the collapsed mass it contains. This can be a one-to-one relation or a distribution.
    \item The probability of a region of volume $V_{0}$ to be ionized $P_{\text{ion}}(V_{0})$ is then computed differently depending on the case : 
    \begin{itemize}
        \item Using Equation \ref{Eq:Pion_NoDispersion} if the $M_{coll}(V_{\text{ion}})$ is one-to-one
        \item If a dispersion is assumed, the functional form of its variance has to be defined. We then access $P_{\text{ion}}(V_{0})$ by marginalizing over the collapsed mass distribution (now approximated by its mean and variance) using Equation \ref{Eq:Pion_Dispersion}.
    \end{itemize}
    \item Finally the number per unit volume of ionized bubbles with volume between $V$ and $V+dV$ is given by Equation \ref{Eq:our_BSD}.
\end{enumerate}

\section{Testing the model for parameter inference}\label{Sec:Results}
\subsection{Physical relation between the collapsed mass inside an ionized region and its volume}

	To compare our model with the numerical results of the HIRRAH-21 simulation, we select two functional forms to parameterize the relation between the volume of an ionized region and its collapsed mass content. A simple physical insight would lead us to consider $M_{coll}\propto V_{\text{ion}}$: the ionized volume is proportional to the number of emitted photons, itself proportional to the collapsed mass assuming a linear mass-to-luminosity relation. Nevertheless, we will consider a more general $M_{coll}\propto V_{\text{ion}}^\alpha$ relation for now, a liberty that will be withdrawn later.
	
\subsubsection{Power law relation}\label{sSec:AffineRelation}

	The first and most straightforward choice is to assume a power law relation which can be expressed as :
	
\begin{equation}
	M_{coll}=A V_{\text{ion}}^{\alpha}
\label{Eq:MofR_affine}
\end{equation}
	\noindent
	where $\alpha$ and $A$ are two parameters to be fitted. Though convenient, this first parameterization will not be able to handle the effect of the mass resolution and should be limited to a range of volumes which is not affected by this limit, that is volume $V_{\text{ion}}$ such that $V_{\text{ion}}>\left(\frac{M_{min}}{A}\right)^{1/\alpha}$ where $M_{min}=4$\ODG{9} M$_\odot$ in our case.
	
\subsubsection{Logarithmic softplus}\label{sSec:SoftPlusRelation}
	
	To better fit simulations and observations, we want to be able to account for the effect of the lower mass limit for star forming halos, that exists physically, although at a lower mass than our simulation mass resolution limit. This physical limit is set by the temperature floor that can be reached by hydrogen atomic or molecular cooling. This suggests another parameterization that will be able to account for this change of regime (at the cost of a higher number of free parameters).
	
	\red{In HIRRAH-21, the mass limit $M_{min}$ is the result of the resolution limit. It induces a sharp threshold in mass and so a flat $M_{coll}(V_{ion})$ relation at low volumes. This is very much model dependent: e.g. \citep{Cohen2016} consider three regimes of star formation efficiency depending on the halo mass. For our simulation,} a possible choice for a function that must be flat in a given range (as the collapsed mass cannot be smaller than $M_{min}$) and linear above a given threshold is the Rectifier Linear function, better known as ReLu in the artificial intelligence domain and defined as $ y:x \mapsto \alpha\min (x-x_{th},0)+y_{th}$ with $\alpha$, $x_{th}$, and $y_{th}$  three free parameters and, in our case $y\equiv \log_{10}(R)$ and $x\equiv \log_{10}(M_{coll})$. However this function is piece-wise linear and numerical computation around the breakpoint tends to result in a brutal and artificial drop in the amplitude of the output BSD. For this reason we chose to implement a variation of this function without any breakpoints: the logarithmic Softplus.
	
\begin{equation}
	\log_{10}\left(\frac{M_{coll}}{M_{coll,th}}\right) =\alpha\frac{\ln\left( 1+\exp\left[k\log_{10}\left(\frac{V_{\text{ion}}}{V_{\text{th}}}\right)\right]\right)}{k}
\label{Eq:MofR_SoftPlus}
\end{equation}
\noindent
 where $\alpha$ is the slope of the relation at high volume, $k=10$ is the smoothness of the transition from one regime to the other and $M_{coll,th}$ and $V_{\text{th}}$ are the collapsed mass and corresponding volume at the transition from the regime dominated by the mass limit to a regime where it is negligible. We set $k=10$, as the sharpness of the transition is not critical for this proof-of-concept study, meaning that this parameterization effectively relies on three parameters.

\begin{table*}[t]
    \begin{center}
	\begin{tabular}{c | l l | l l}
	\hline
	\hline
	& \multicolumn{2}{c |}{$\mathbf{\Xhii\sim 1\%}$} & \multicolumn{2}{c}{$\mathbf{\Xhii\sim 3\%}$} \\
	\textbf{Case} & $\errBSD$ & $\errMofV$ & $\errBSD$ & $\errMofV$ \\
    \hline
    Power law & 0.033 & 0.020 & 0.12 & 0.039 \\
    Power law Poisson & 0.047 & 0.053 & 0.17 & 0.059 \\
    Power law Dispersion & 0.063 & 0.13 & 0.33 & 0.31 \\
    Softplus & 0.036 & 0.010 & 0.079 & 0.025 \\
    Softplus Poisson & 0.029 & 0.044 & 0.083 & 0.045 \\
    Softplus Dispersion & 0.058  & 0.00026 & 0.077 & 0.0013 \\
	\hline
	\end{tabular}
	\caption{Mean squared of the BSD fitting process ($\errBSD$) and of the physical $M_{coll}(V_{\text{ion}})$ relation reconstruction ($\errMofV$) for all the models discussed in Sections \ref{sSec:ParamInference} and \ref{sSec:BackwardDisp} at two global ionization fraction $\Xhii\sim 1\%$ and $\Xhii\sim 3\%$. }
\label{Table:Khi}
\end{center}
\end{table*}

\subsection{Parameter inference}\label{sSec:ParamInference}

	One key application of the model developed in this study is that it can be used to infer the relation $M_{coll}(V_{\text{ion}})$ from the BSD and thus to constrain the underlying astrophysical processes. For that, one has to assume a parametric form for the relation and apply optimization algorithms like gradient-descent or Markov-Chain Monte-Carlo (MCMC) to fit a given observed  BSD. For this study, we apply a simple grid-search method to illustrate the capabilities of this model. \red{To quantify the error we will use a mean squared error on the logarithm of the quantities which can be written:
	\begin{equation}
	\err (x) = \frac{1}{n}\sum_{i=1}^{n} \left( y_{i} - y_{i}^{\text{true}}\right)^{2}
	\label{Eq:MSElog}
	\end{equation}
	where $y^{\text{true}}$ is the reference value. In our case, $y^{\text{true}}$ is always the numerical value computed from HIRRAH-21 simulation.} We first focus on low global ionization fractions ($\Xhii\sim 1\%$ and $\Xhii\sim 3\%$) where the effect of percolation is negligible.\\
	
\begin{figure*}[t]
\begin{center}
\includegraphics[width=\columnwidth]{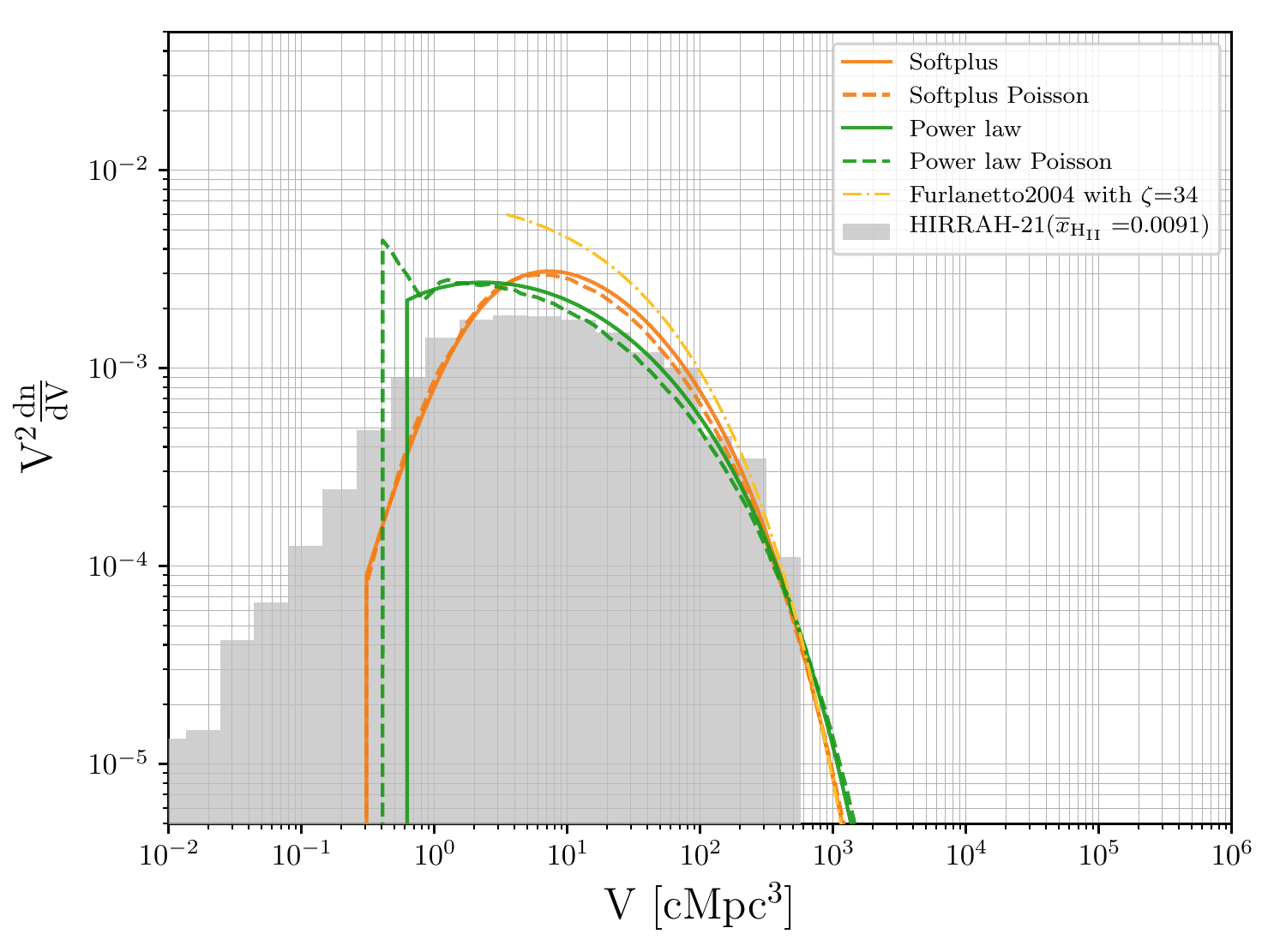}
\includegraphics[width=\columnwidth]{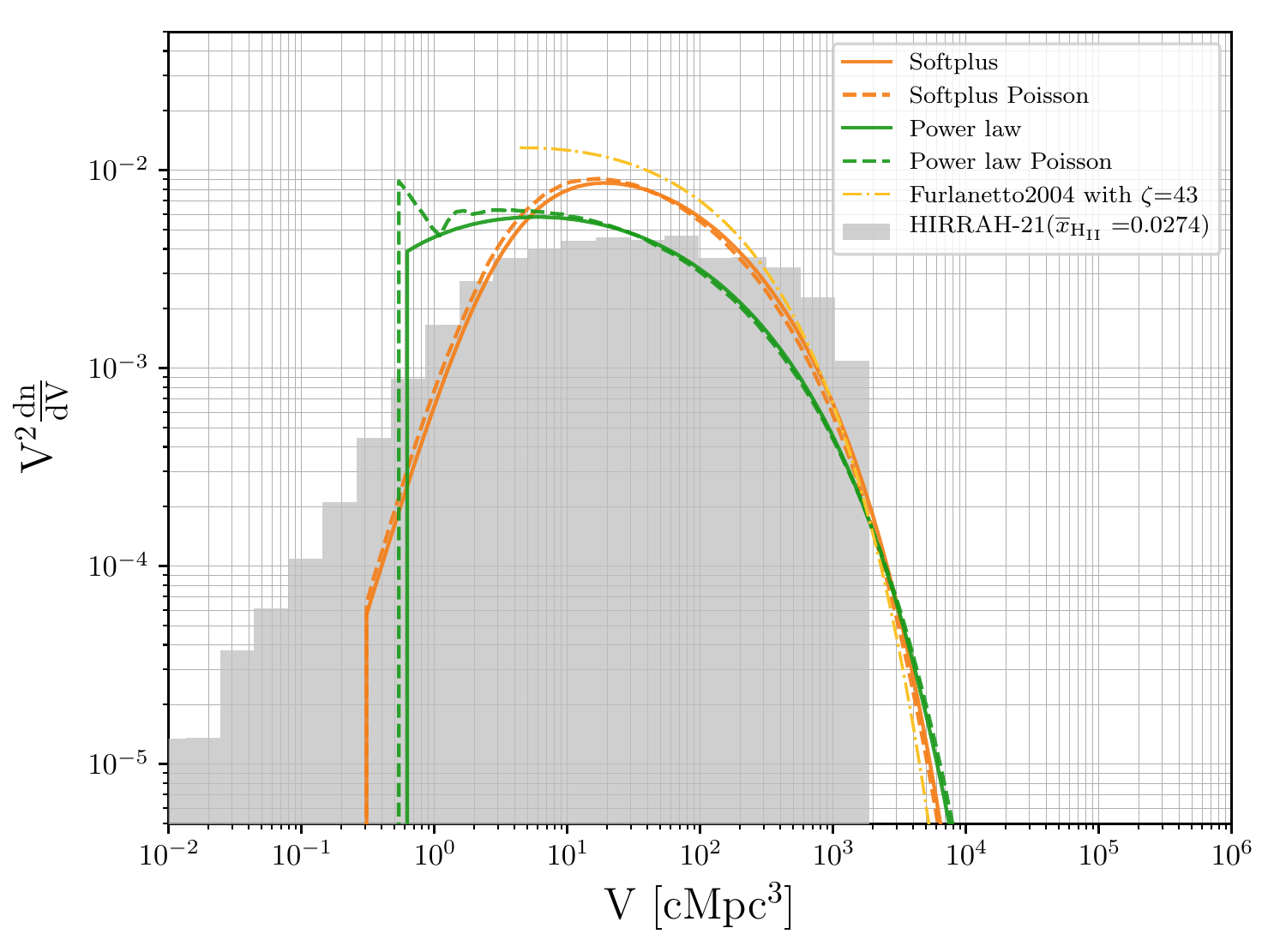}
\caption{Various fit of the reference numerical BSD (gray histogram) using a power law parameterization from Section \ref{sSec:AffineRelation} (green lines) or a logarithmic Softplus parameterization from Section \ref{sSec:SoftPlusRelation} (orange lines) considering the sample variance (dashed lines) or not (solid lines). A numerical fit of the reference BSD from \citet{Furlanetto2004a} (Eq.\ref{Eq:BSD_Furlanetto}) using only the efficiency parameter $\zeta$ is shown as a comparison (yellow dashed line). The fit is done at $\Xhii\sim 1\%$ (left panel) and $\Xhii\sim 3\%$ (right panel).}
\label{Fig:BackwardBSD}
\end{center}
\end{figure*}

	In Fig.\ref{Fig:BackwardBSD} we use both of our power law (green lines) and Softplus (orange lines) parameterization, with (dashed lines) and without (solid lines) sample variance modelled by the Poisson distribution. We fit the BSD from the HIRRAH-21 simulation (gray histogram) to infer the  parameters. The method is applied at two different global ionization fraction of $\Xhii\sim 1\%$ (left panel) and $\Xhii\sim 3\%$ (right panel). We also show the best-fit FHZ04 model for reference. We see that all our best fits are reasonably close to the numerical distribution. Table \ref{Table:Khi} presents the resulting mean squared errors for the BSD best-fits $\errBSD$ and of the corresponding reconstructed $M_{coll}(V_{\text{ion}})$ relations compared to the numerical relation $\errMofV$, for both $\Xhii\sim 1\%$ and $\Xhii\sim 3\%$. For all the current models, it shows that the fit mean squared error is, at worst, $17\%$ but always under $10\%$ for the Softplus parameterization.
	
	However the theoretical BSD seems to overestimate the total ionization fraction as illustrated by their respective maximums that have an amplitude $\sim50\%$ greater than the numerical one. As expected, the decreasing population of ionized bubbles with decreasing volume due to the mass resolution limit is better fitted when using the Softplus model and the overall shape appears to be closer to the numerical reference. Our model also gives results closer to the numerical distribution for all cases than the FHZ04 model, especially at volumes $V\lesssim 10^{2}$cMpc$^{3}$.

\begin{figure*}[t]
\begin{center}
\includegraphics[width=\columnwidth]{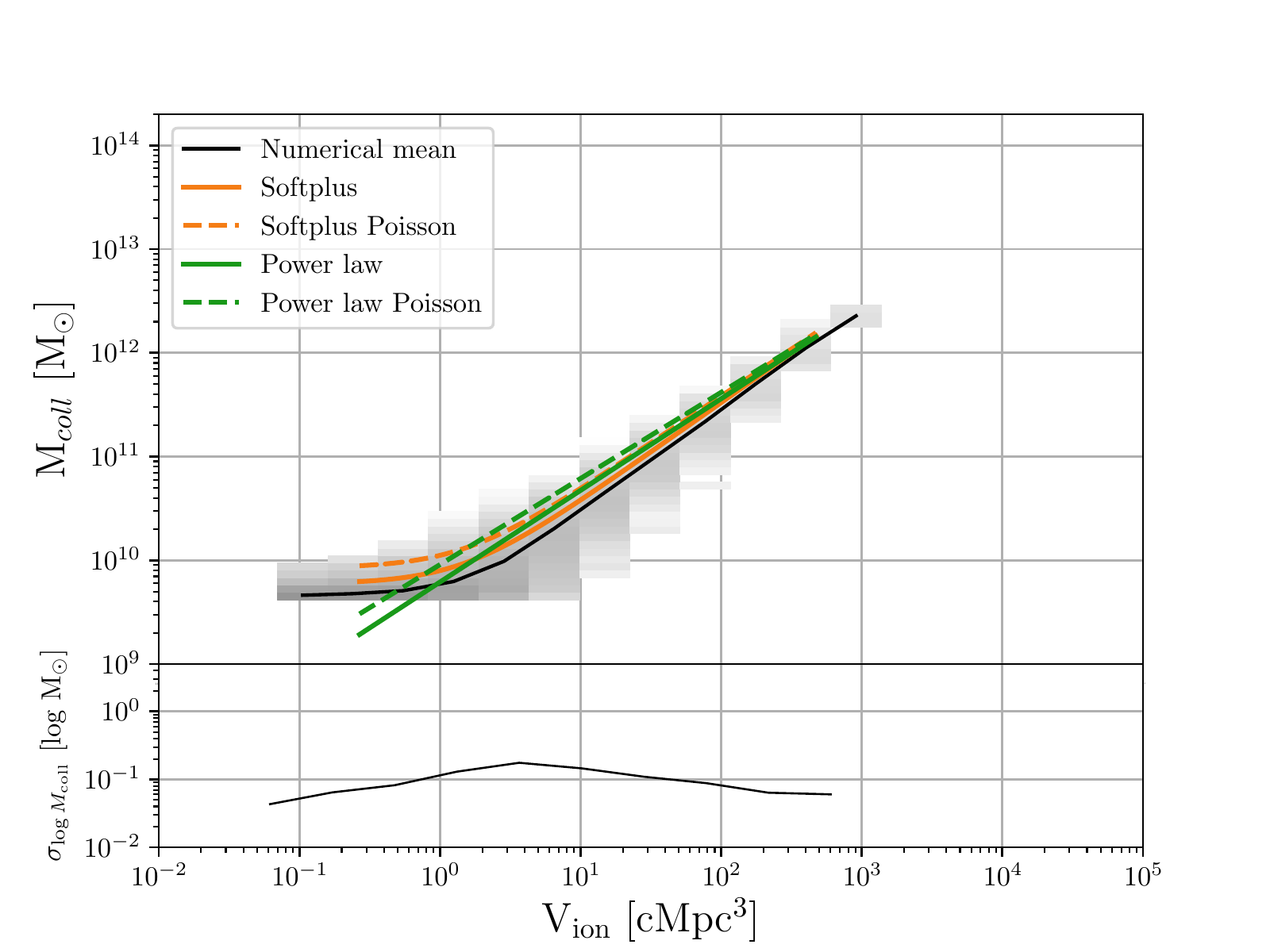}
\includegraphics[width=\columnwidth]{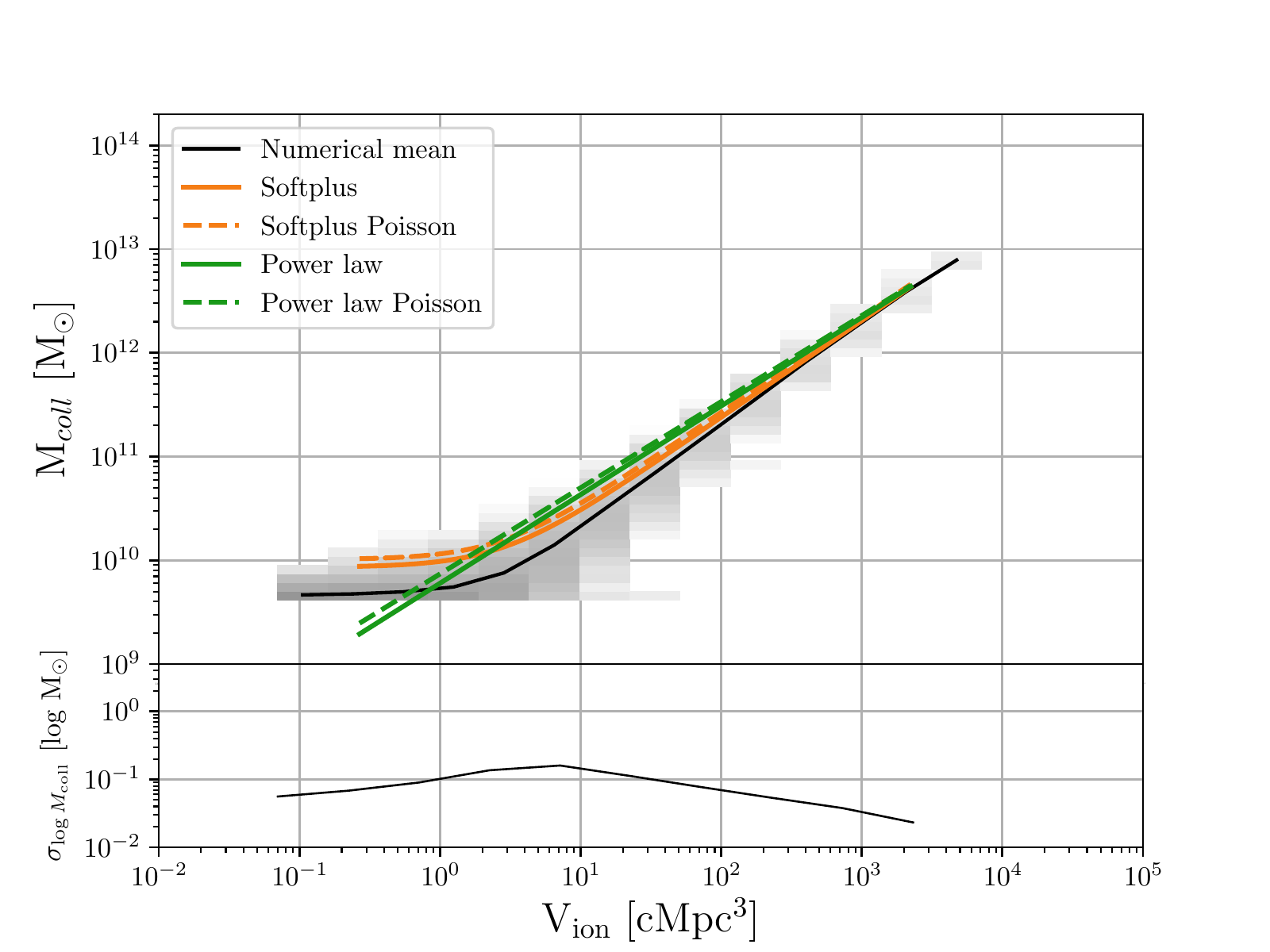}
\caption{Collapsed mass inside an ionized region as a function of its volume. The gray colormap represents the distribution of halos the HIRRAH-21 simulation and its numerical mean (top panel) and dispersion (bottom panel) are shown as black lines in the corresponding panels. The inferred relations with (dashed lines) and without (solid lines) considering the sample variance are shown for both our power law (green lines) and our logarithmic Softplus (orange lines) parameterizations. The fit is done at $\Xhii\sim 1\%$ (left panel) and $\Xhii\sim 3\%$ (right panel).}
\label{Fig:Backward_M_of_V}
\end{center}
\end{figure*}

	Fitting the relatively smooth shape of the numerical BSD with 2 and 3 parameters models is not in itself a strong result. That the parameters and the model are based on a physical understanding of the system is more interesting. But showing that the inferred values for the parameters yield a $M_{coll}(V_{\text{ion}})$ relation that actually matches what we can directly compute from simulations data would tell us that when we apply this procedure to an observed BSD, we will learn something relevant about the process of reionization. We show in Fig.\ref{Fig:Backward_M_of_V}, for both $\Xhii\sim 1\%$ (left panel) and $\Xhii\sim 3\%$ (right panel), the inferred physical relation $M_{coll}(V_{\text{ion}})$ for the power law (green lines) and Softplus (orange lines) parameterizations, with (dashed lines) and without (solid lines) considering the sample variance, and the distribution for the same relation computed directly with the data from the simulation. We see that all parameterizations infer relatively reasonable physical relations. However all our inferred relations overestimate the average amount of collapsed mass inside a region of a given volume, especially for the smallest volumes ($V\lesssim 2$\ODG{1}cMpc$^{3}$), compared to the average value directly computed from the simulation. Referring to Table \ref{Table:Khi} it leads to a $M_{coll}(V_{\text{ion}})$ mean squared error of $\sim 5\%$ for models considering the sample variance and $\sim 2.5\%$ for the others.
	
	This difference must be tempered by the possible uncertainties in both the simulation and the model results. Concerning HIRRAH-21 simulation, a Friend-of-Friend algorithm is run on the particle distribution to construct halos. This halo finder especially relies on a linking length that roughly quantify the maximal distance for two particles to be considered as neighbours inside the same halo \citep{Davis1985}. Various values of the linking length can be considered reasonable and choosing one rather than another impacts the resulting $M_{coll}(V_{\text{ion}})$ relation. For the analytical model, it has been specifically designed to be adjustable and we show in Fig.\ref{Fig:Pmcoll_Comp} that the collapsed mass probability already differs by a factor 2 between the two depicted halo mass function formalisms. The inferred $M_{coll}(V_{\text{ion}})$ can therefore vary for two different CMFs.

	In Fig.\ref{Fig:BackwardBSD}, we see that the difference induced by including the sample variance is relatively negligible in our case. In term of physical relation inference, Fig.\ref{Fig:Backward_M_of_V} and Table \ref{Table:Khi} shows that models including sample variance perform worse than the others, being less in agreement with the numerical mean relation. It is possible that the sharp mass cutoff considered in the Poisson distribution approach is a poor description of the actual mass resolution limit in the simulation.  Consequently, although it is theoretically more comprehensive to account for this effect,  we will from now on only look at cases where the sample variance is not considered. \red{Furthermore, for all cases the slope $\alpha$ of our mean relation in Equations \ref{Eq:MofR_affine} and \ref{Eq:MofR_SoftPlus} is always close to the expected value of $1$. From now on we will fix it at this value, effectively withdrawing one free parameter from both mean relations.}

\section{Injecting dispersion in the $M_{coll}(V_{ion})$ relation}

	In Section \ref{Sec:Results}, we showed that for both of our parameterizations, our best fits BSD have a moderate deficit of large bubbles and excess of mid-sized bubbles. This problem could be partially solved if we consider a dispersion in the physical $M_{coll}(V_{\text{ion}})$ relation. In our numerical simulation this dispersion does exist and is non-negligible, as can be seen in Fig.\ref{Fig:Backward_M_of_V}. The effect of this dispersion could be important considering that a variation of the mean $M_{coll}$ results in a strong variation in term of collapsed mass probability distribution as we can see in Fig.\ref{Fig:Pmcoll_Comp}.

\subsection{Dispersion parameterizations}

	We already detailed, in Section \ref{sSec:Th_AddDispersion}, the required modification to the model in order to include dispersion in the $M_{coll}(V_{\text{ion}})$ relation. We now need to choose a parameterization. As shown in Fig.\ref{Fig:Backward_M_of_V}, the dependence of the standard deviation as a function of $V_{\text{ion}}$ exhibits two different trends depending on the dominance of the mass resolution effect. However, the slopes are not sharp and to avoid adding too many free parameters, we decide to use the simple parameterization:

\begin{equation}
	\frac{\sigma \left( M_{coll}\right)}{M_{coll}} =B V_{\text{ion}}^{\beta}
\label{Eq:dispMofR_affine}
\end{equation}
\noindent
	with $B$ and $\beta$ as our free parameters. \red{Since we set $\alpha$ to $1$ in the mean relation,}  using Equation \ref{Eq:dispMofR_affine} along with the mean relations of Equation \ref{Eq:MofR_affine} we will be fitting a total of three free parameters while with Equation \ref{Eq:MofR_SoftPlus} we will be using four free parameters in total.

\subsection{Fitting the bubble size distribution with models including dispersion}\label{sSec:BackwardDisp}

\begin{figure*}[t]
\begin{center}
\includegraphics[width=\columnwidth]{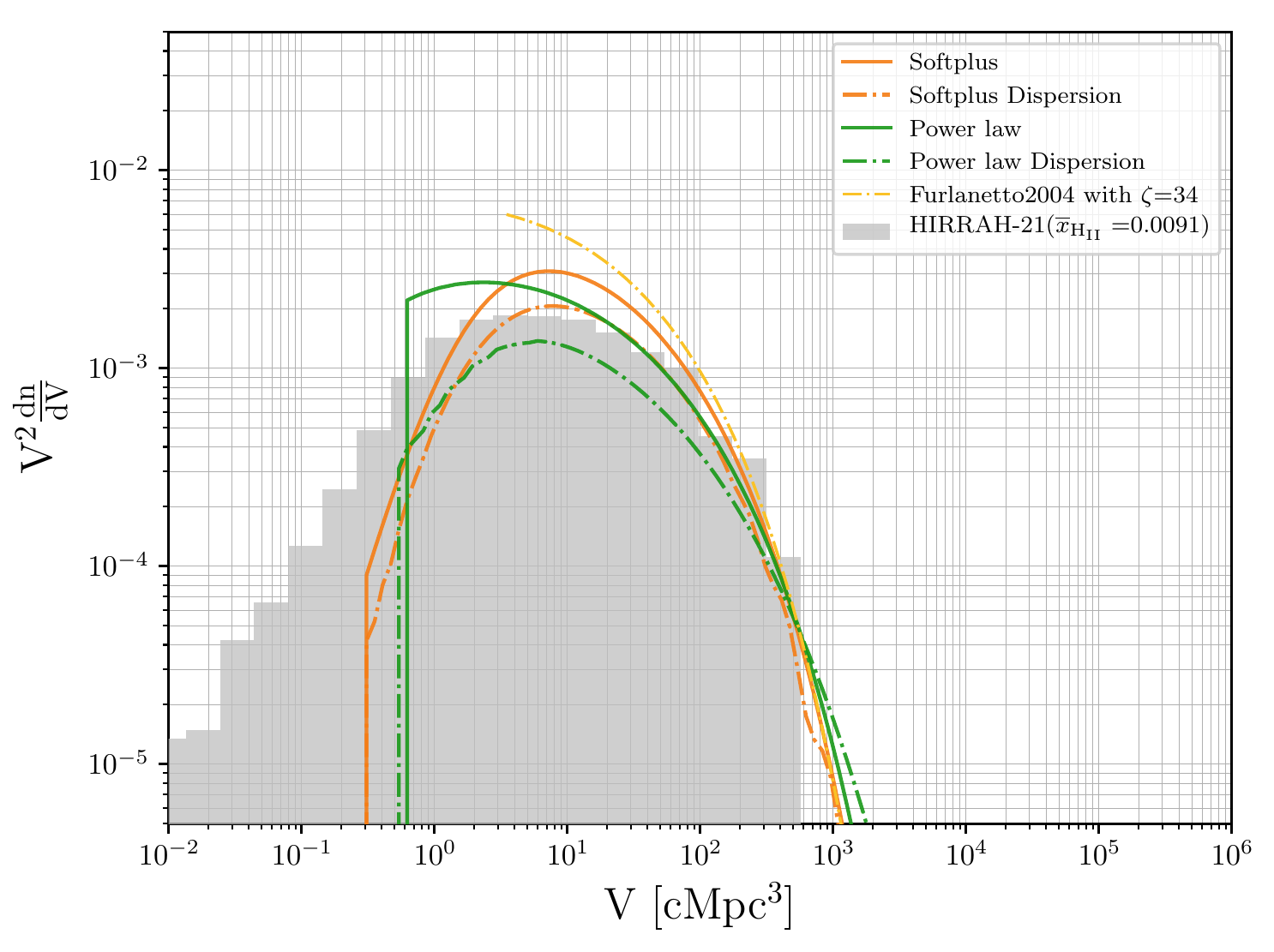}
\includegraphics[width=\columnwidth]{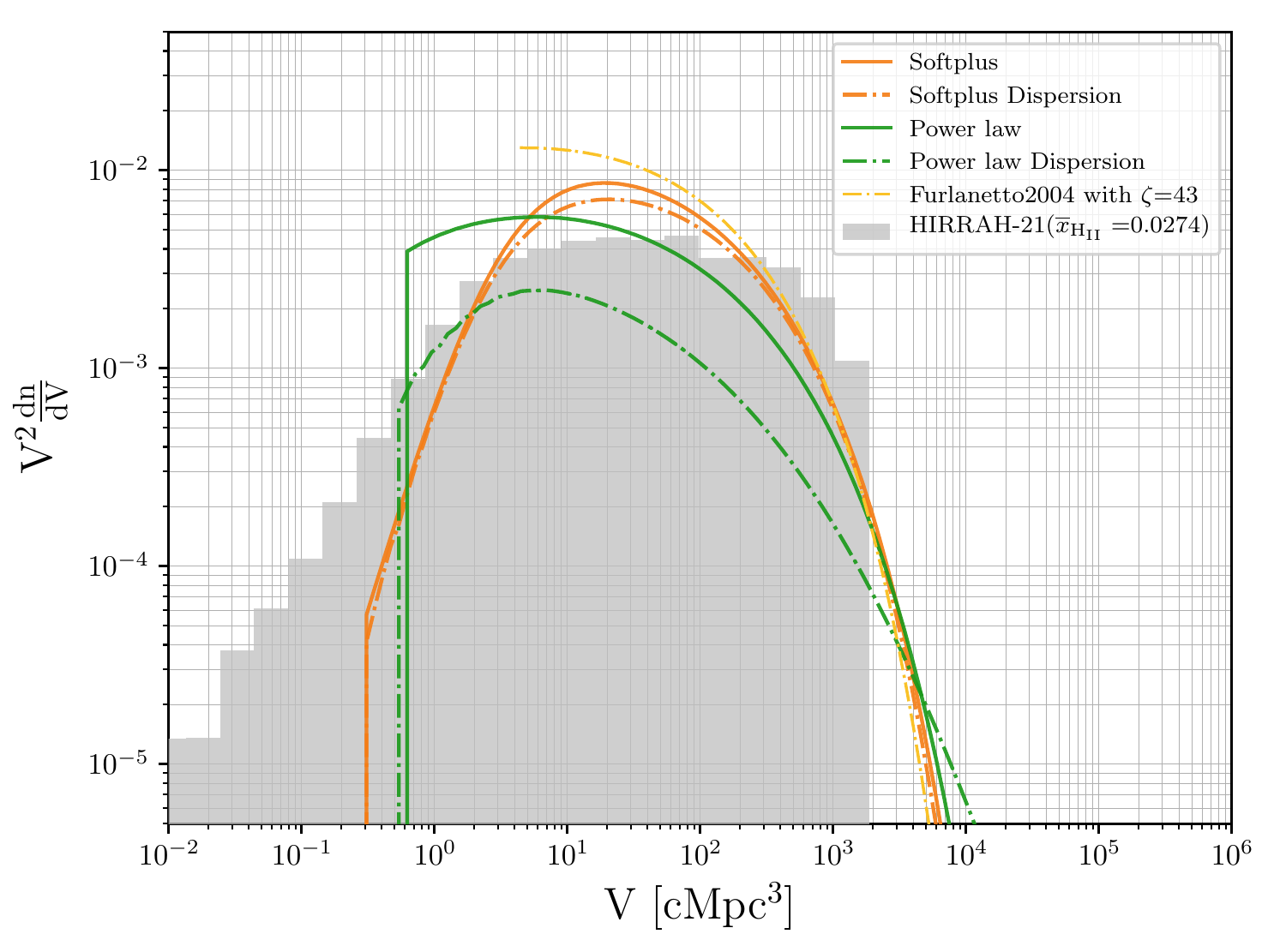}
\caption{Various fit of the reference numerical BSD (gray histogram) using power law parameterization from Section \ref{sSec:AffineRelation} (green lines) or logarithmic Softplus parameterization from Section \ref{sSec:SoftPlusRelation} (orange lines) considering the dispersion (dashed lines) or not (solid lines). A numerical fit of the reference BSD using Equation \ref{Eq:BSD_Furlanetto} from \citet{Furlanetto2004a} inferring the efficiency parameter $\zeta$ is shown as a comparison (yellow dashed line). The fit is done at $\Xhii\sim 1\%$ (left panel) and $\Xhii\sim 3\%$ (right panel).}
\label{Fig:BackwardBSD-Disp}
\end{center}
\end{figure*}

	In Fig.\ref{Fig:BackwardBSD-Disp} we use both of our parameterizations detailed in Sections \ref{sSec:AffineRelation} (green lines) and \ref{sSec:SoftPlusRelation} (orange lines) with (dashed lines) and without (solid lines) considering dispersion. We fit the BSD from the HIRRAH-21 simulation (gray histogram) to infer the underlying parameters of this parameterizations. The method is also applied at two different global ionization fractions of $\Xhii\sim 1\%$ (left panel) and $\Xhii\sim 3\%$ (right panel). For both parameterizations, considering the dispersion results in a decrease of the global amplitude of the theoretical best-fit BSD. On the whole range however, the fit to the numerical BSD is not really improved when considering dispersion.

\begin{figure*}[t]
\begin{center}
\includegraphics[width=\columnwidth]{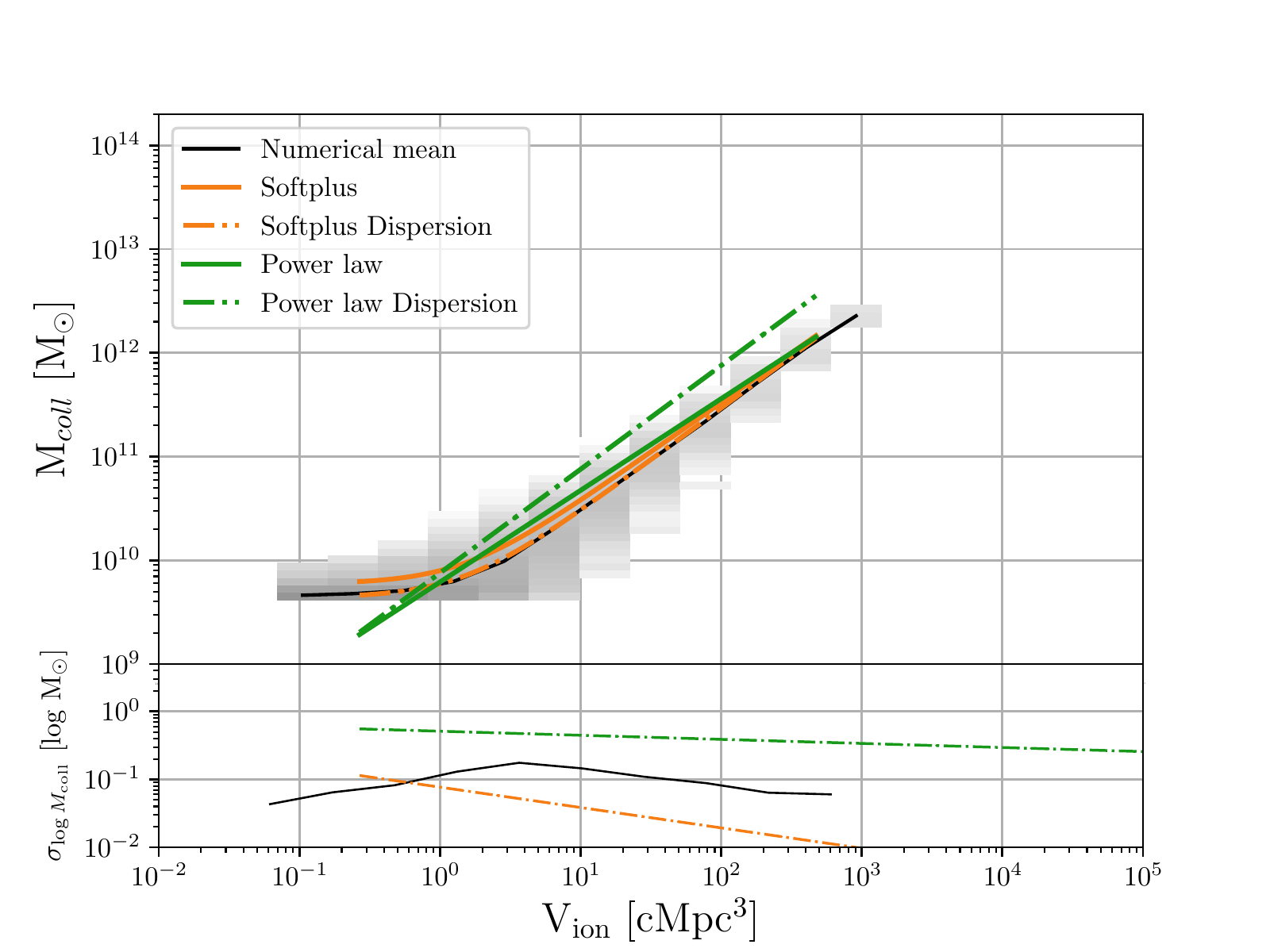}
\includegraphics[width=\columnwidth]{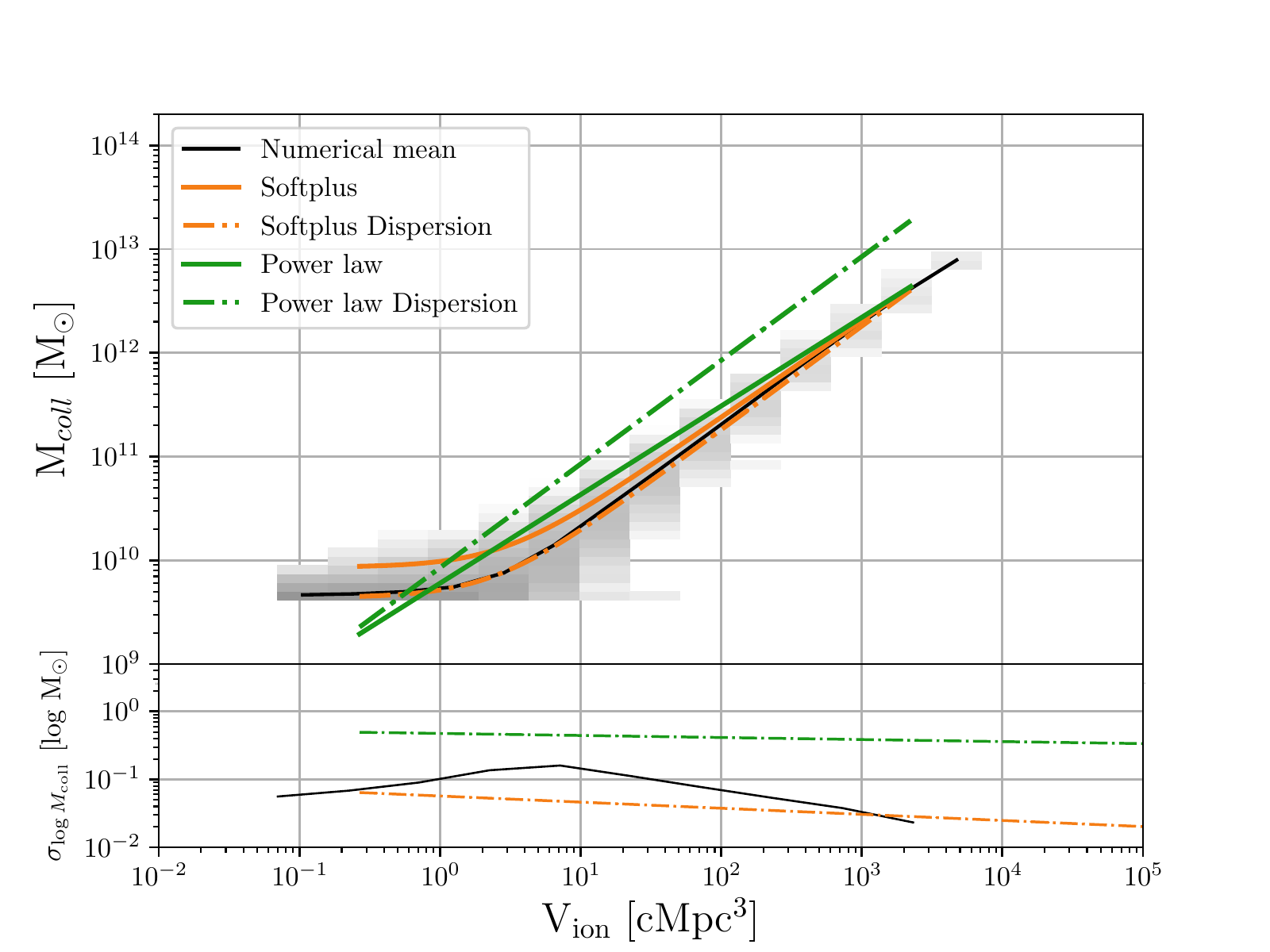}
\caption{Collapsed mass inside an ionized region as a function of its volume. The gray colormap represents the distribution of halos the HIRRAH-21 simulation and its numerical mean and dispersion are shown as black lines in the corresponding panels. The inferred relations with (dashed lines) and without (solid lines) considering the dispersion are shown for both our power law (green lines) and our logarithmic Softplus (orange lines) parameterizations. The fit is done at $\Xhii\sim 1\%$ (left panel) and $\Xhii\sim 3\%$ (right panel).}
\label{Fig:Backward_M_of_V-Disp}
\end{center}
\end{figure*}

	In Fig.\ref{Fig:Backward_M_of_V-Disp} we show the inferred $M_{coll}(V_{\text{ion}})$ relation with (dashed lines) and without (solid lines) including the dispersion for both our power law (green lines) and our logarithmic Softplus (orange lines) parameterizations. The cases with dispersion highlight the fact that the inference capability of our model strongly depends on the choice of an adequate functional form.\\
	
	Here, the simple power law model being less flexible and unreasonable for volumes where the mass resolution effect is dominant (V$\lesssim 2$cMpc$^{3}$), the resulting theoretical BSD is unable to match the numerical BSD. To be as close as possible to the numerical BSD within the constraints  of this functional form, the best-fit model overestimates the dispersion amplitude by a factor $\sim 3$. On the contrary, when using the logarithmic Softplus relation of Equation \ref{Eq:MofR_SoftPlus}, a functional form that is flexible enough to account for the mass resolution effect, the inferred mean $M_{coll}(V_{\text{ion}})$ relation is in very good agreement with the numerical one, though the dispersion seems to be somewhat underestimated in this case. In Table \ref{Table:Khi} we see that using the logarithmic Softplus relation in case of dispersion leads to a mean squared error $\lesssim 0.1\%$ on the logarithm of the inferred physical relation $M_{coll}(V_{\text{ion}})$ for both $\Xhii\sim 1\%$ and $\Xhii\sim 3\%$.

\subsection{Model sensitivity to the halo mass formalism}\label{sSec:CMF_sensitivity}

	The previous analysis was performed using the conditional mass function of Equation \ref{Eq:CMF_ST} but, as stated in Section \ref{sSec:Theory_CMF} our model can use various functional forms for the conditional mass functions which will logically affect the resulting BSD.

\begin{figure}[t]
\begin{center}
\includegraphics[width=\columnwidth]{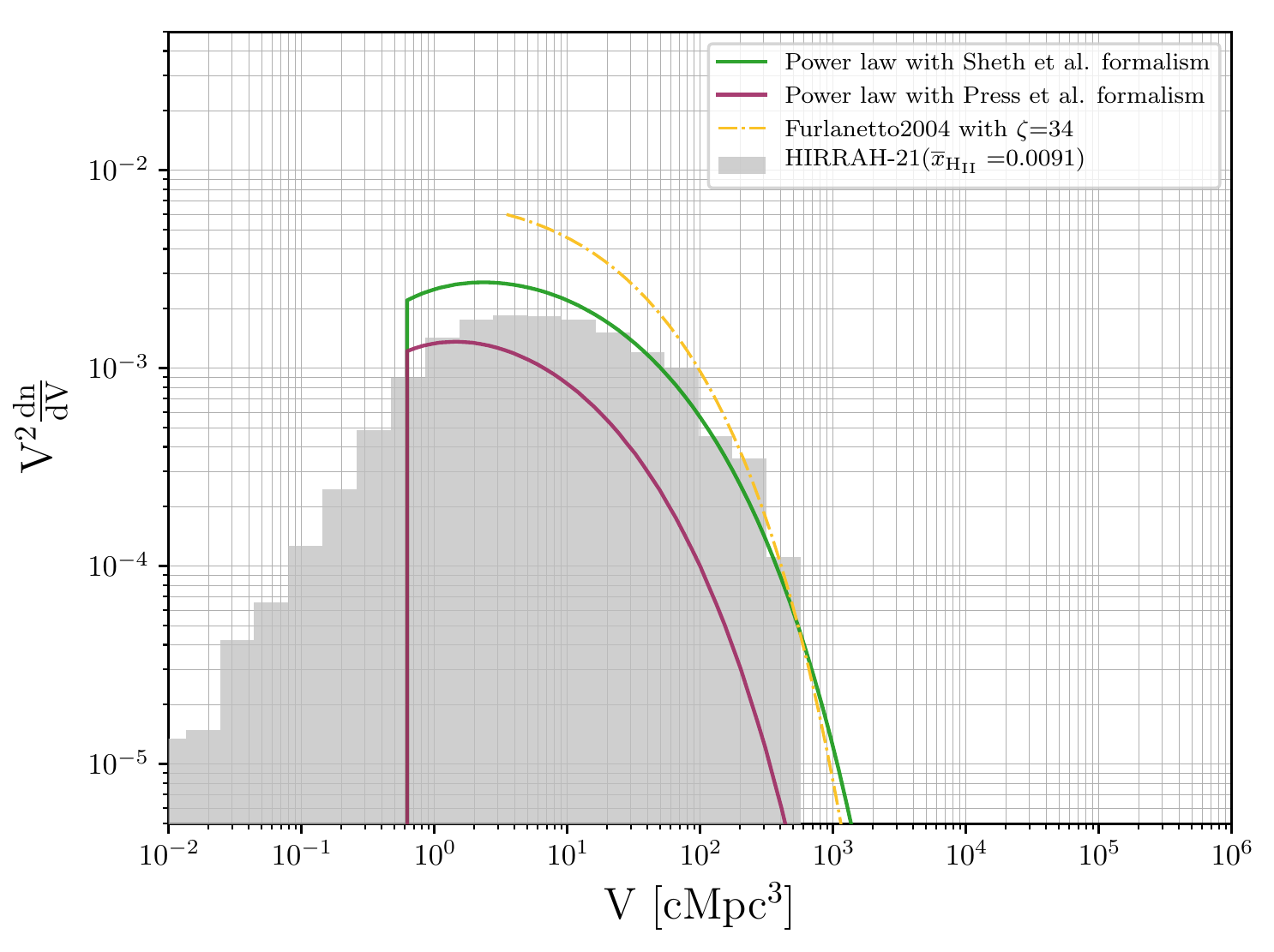}
\caption{Bubble size distribution at $\Xhii\sim 1\%$ using the same parameters (inferred with the model using \citet{Sheth1999}) in Equation \ref{Eq:MofR_affine} using the CMF of Equation \ref{Eq:CMF_ST} based on \citet{Sheth1999} (green line) and the CMF of Equation \ref{Eq:CMF_EPS} based on \citet{Press1974} (purple line). The Numerical BSD from the HIRRAH-21 simulation and a numerical fit of the reference BSD using Equation \ref{Eq:BSD_Furlanetto} from \citet{Furlanetto2004a} inferring the efficiency parameter $\zeta$ (yellow line) are shown for comparison.}
\label{Fig:CMF_discrepancy}
\end{center}
\end{figure}

In Fig.\ref{Fig:CMF_discrepancy} we represent the resulting BSD at $\Xhii\sim 1\%$ for the same power law relation between the collapsed mass inside an ionized region and its volume using the CMF of Equation \ref{Eq:CMF_ST} based on \citet{Sheth1999} (green line) and the CMF of Equation \ref{Eq:CMF_EPS} based on \citet{Press1974} (purple line). We observe an uniform decrease of the amplitude on the whole range for the CMF based on \citet{Press1974} compared to the one based on \citet{Sheth1999}. The amplitude decrease can be explained by the fact that, for a given volume, the region that are ionized are most likely the ones with a large collapsed fraction. In Fig.\ref{Fig:Pmcoll_Comp} we showed that the collapsed mass probability distribution peak is shifted to lower collapsed mass for \citet{Press1974} compared to  \citet{Sheth1999}. Therefore, a given volume is less likely to have a high collapsed mass for the former than for the latter, which consequently leads to an overall lower occurrence of bubbles in \citet{Press1974} case for the same physical relation.\\

	It is worth noting that the previous result is not a performance comparison between the two CMFs. Indeed the parameter values for the physical relation $M_{coll}(V_{\text{ion}})$ are those from the best fit using the CMF of \citet{Sheth1999}. The theoretical BSD using \citet{Sheth1999} has thus been fitted to match the numerical BSD while the BSD using \citet{Press1974} CMF has not. This comparison strongly emphasizes the fact that the choice of mass function formalism has an effect which is non-negligible.

\section{Extending the model in the percolation regime}\label{Sec:Percolation}

In the previous section, we showed that, granted an adequate functional form, the inference capability of our model is especially high when considering dispersion. However, even using dispersion, the inference power and even the BSD matching capability of our model strongly degrades for $\Xhii > 10\%$. This effectively limits its inference possibilities to the early stages of the reionization process. The degradation is mainly due to the well-known percolation process \citep{Furlanetto2016}, that has to be implemented in our model.\\

\subsection{Implementing the percolation effect}

    To account for the percolation process without extensively modifying our model, we implement an algorithm that empirically emulates the effect of percolation on a distribution of bubbles where the possibility of overlap has been ignored. It thus takes as an input the BSD $V^{2}\frac{dn_{b}}{dV}$ of our model. The algorithm acts on each bin of volume $V_{i}$, starting from the lowest, and proceeds as follows :
	
\begin{enumerate}
\item First, we assume random, uncorrelated locations for bubble centers. If the center of a bubble of a given volume $V_{j}>V_{i}$ is closer than $R_{i}+R_{j}$ (where $R_{x}$ is the radius corresponding to $V_{x}$) to the center of a bubble of volume $V_{i}$, they are overlapping. There is therefore a shell around each bubble of volume $V_{j}$ that generates an overlap. Considering all bubbles of size $V_{j}$, the total volume, per unit volume, where overlap with bubble of size $V_{i}$ can happen is : 
$$V_{\text{over}}^{i|j}= f_{f}\left[\frac{4\pi}{3}(R_{i}+R_{j})^{3} - V_{j} \right] dn_{b}(V_{j})$$ 
where $dn_{b}(V_{j})$ is the comoving number density of bubbles of volume $V_{j}$ and $f_{f}$ is an empirical fudge factor. This fudge factor is needed to roughly account for the fact that ionized bubbles, like the sources of ionizing photons, actually tend to be clustered and not homogeneously distributed.

\item The number of bubbles of volume $V_{i}$, per unit volume, whose center falls in this overlapping zone (assuming uncorrelated center positions) is: $$dn_{b}^{i|j}= V_{\text{over}}^{i|j}dn_{b}(V_{i})$$ Assuming that each percolation only involves one bubble of volume $V_{i}$ and one of volume $V_{j}$, $dn_{b}^{i|j}$ is the number of bubbles in the two volumes bin that will percolate with a bubble of the other bin.
\item There are $dn_{b}^{i|j}$ bubbles of volume $V_{i}$ and of volume $V_{j}$ that  percolate, thus the comoving number density of ionized bubbles in these two volumes bins is decreased accordingly :
\begin{align*}
	dn_{b}(V_{i}) & =dn_{b}(V_{i}) - dn_{b}^{i|j}\\
	dn_{b}(V_{j}) & =dn_{b}(V_{j}) - dn_{b}^{i|j}
\end{align*}
\item We approximate the volume of the bubbles that result from the percolation process as $V_{i}+V_{j}$ (this is obviously the value in the case where the two bubbles are just touching, the actual average value should be somewhat smaller, but also depends on the relative values of $V_i$ and $V_j$). The comoving number density of ionized bubbles at this volume is thus increased by $$ dn_{b}(V_{i}+V_{j}) = dn_{b}(V_{i}+V_{j}) + dn_{b}^{i|j}$$
\end{enumerate}

 When all volume bins $V_{j}$ have been considered for a given volume $V_{i}$, the algorithm selects the next bin (in increasing volume order) and repeats the previous steps.

\subsection{Results at $\Xhii\gtrsim 10\%$}

\begin{figure}[t]
\begin{center}
\includegraphics[width=\columnwidth]{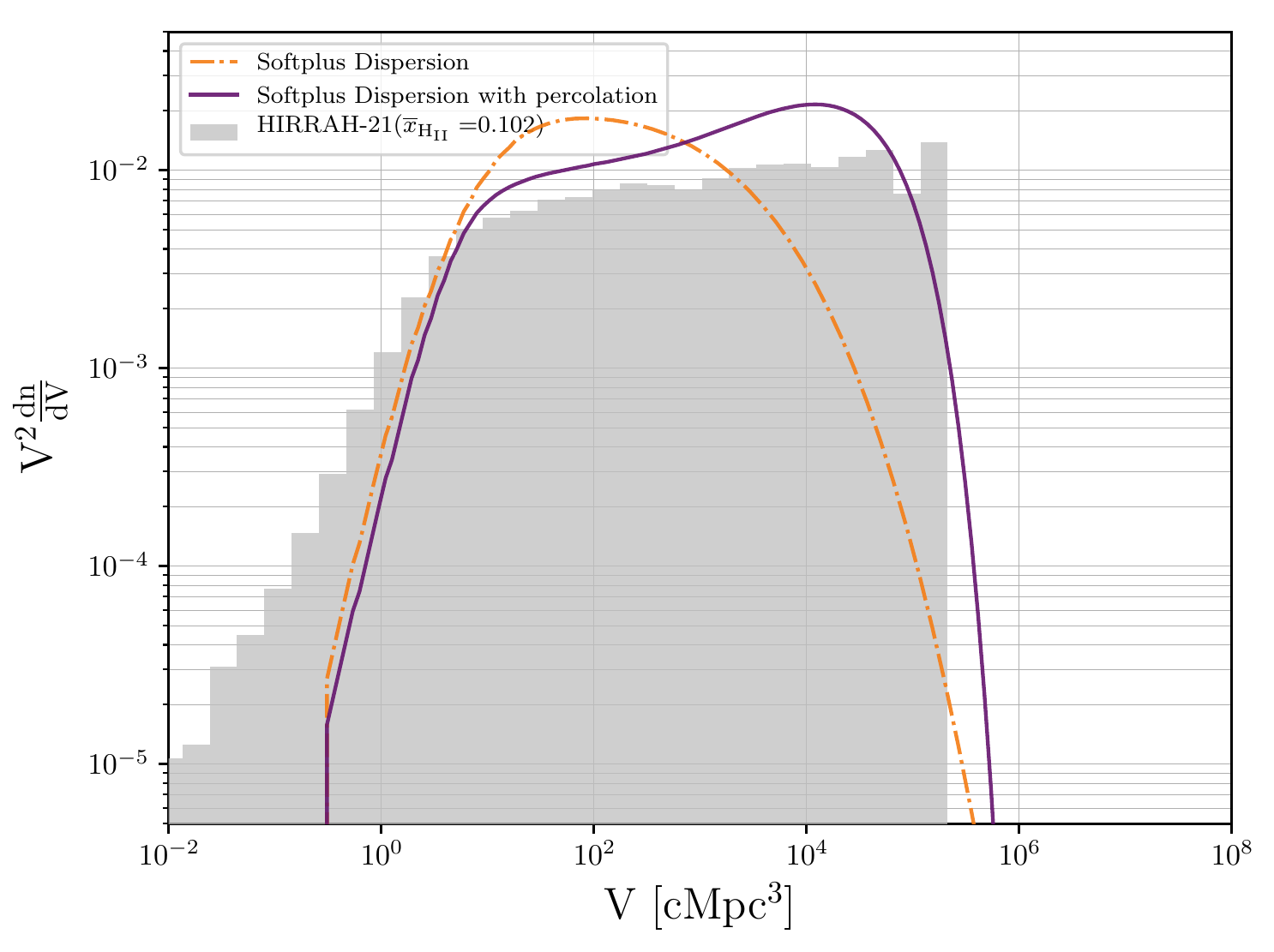}
\includegraphics[width=\columnwidth]{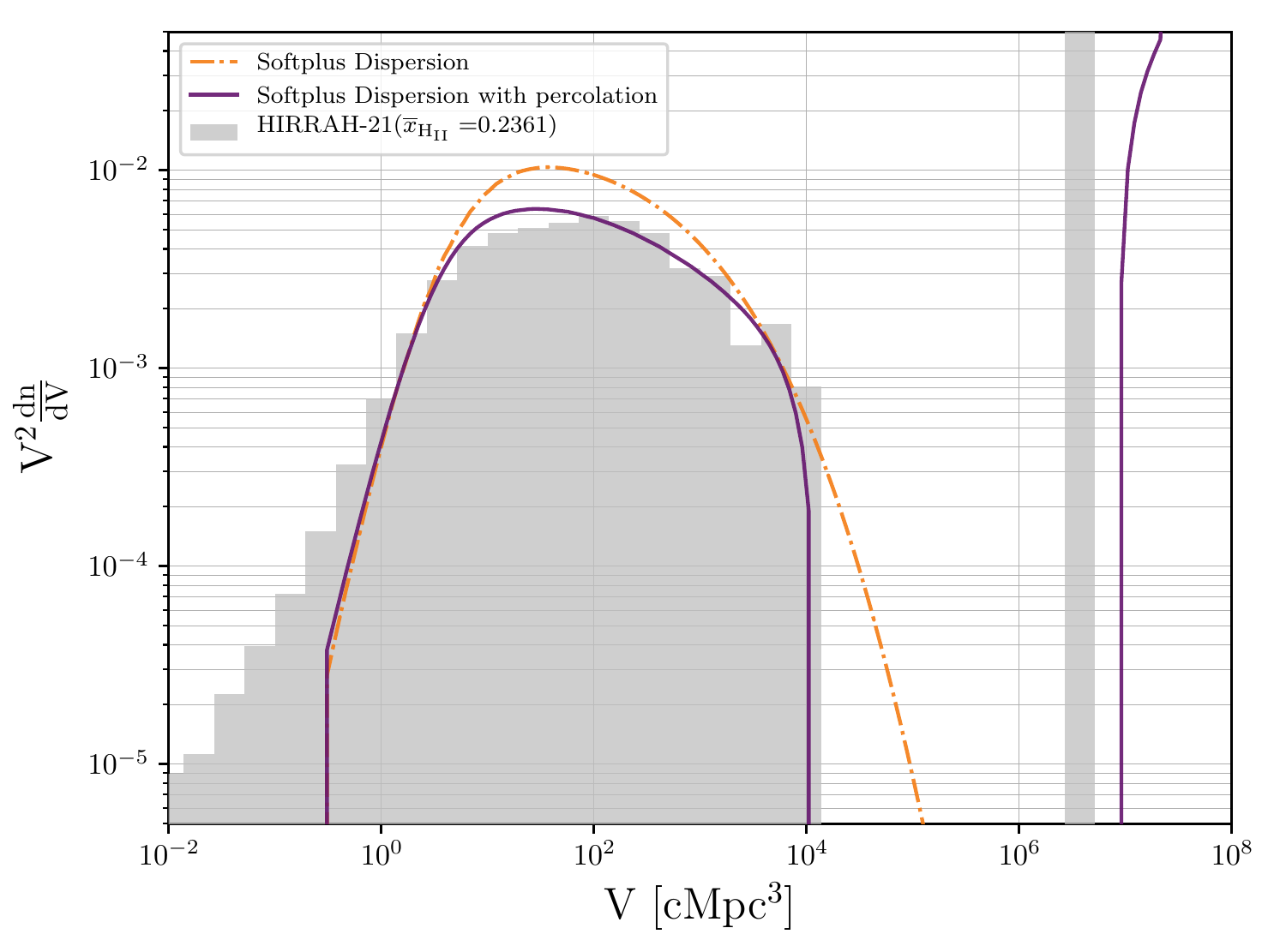}
\caption{Fit of the reference numerical BSD (gray histogram) at $\Xhii\sim 10\%$ (top) and $\Xhii\sim 22\%$ (bottom) using the logarithmic Softplus parameterization from Section \ref{sSec:SoftPlusRelation} considering the dispersion with percolation (purple line) and without percolation (orange line).}
\label{Fig:BSD_percolation}
\end{center}
\end{figure}

\begin{table*}[t]
    \begin{center}
	\begin{tabular}{c | c | c | c | c | c | c}
	\hline
	\hline
	$\err\left(\log_{10}\left[\text{max}\left( V^{2}\frac{dn_{b}}{dV},10^{-5}\right)\right]\right)$ & $\Xhii\sim 10\%$ & $\Xhii\sim 16\%$ & $\Xhii\sim 22\%$ & $\Xhii\sim 30\%$ & $\Xhii\sim 38\%$ & $\Xhii\sim 50\%$ \\
    \hline
    Softplus Dispersion & 0.76 & 0.08 & 0.034 & 0.048 & 0.057 & 0.068\\
    Softplus Dispersion + Percolation & 0.12 & 0.042 & 0.026 & 0.029 & 0.006 & 0.026\\
	\hline
	\end{tabular}
	\caption{Mean squared error computed on $\log_{10}\left[\text{max}\left( V^{2}\frac{dn_{b}}{dV},10^{-5}\right)\right]$ for $5\times 10^{-1}\lesssim V \lesssim 10^{6}$ for the logarithmic Softplus parameterization using dispersion with and without using the percolation algorithm at various global ionization fraction ($\Xhii\sim 10\%$, $16\%$, $22\%$, $30\%$, $38\%$, and $50\%$). }
\label{Table:KhiBSD_percolation}
\end{center}
\end{table*}

		To assess the validity of our percolation algorithm we fit the BSD at $\Xhii\gtrsim 10\%$ where the percolation effect is dominant. Without prior knowledge on the value of the fudge factor $f_{f}$, we consider it as a free parameter to be fitted. In this regime where the percolation is non-negligible, ionized bubbles more voluminous than a given volume have virtually all percolated into a single percolated cluster that spans the whole Universe, the remaining bubble growing inside neutral patches embedded into this percolated cluster. A sharp cut-off is therefore expected at large volume, creating a discontinuity in the BSD and making the fitting process tricky. We decided not to include the percolated cluster in the fit, as its volume depends on the simulation box size in the numerical simulation whereas it tends to infinity in our theoretical percolation algorithm. Additionally, if our fitted model happens to have a sharp cutoff at large volumes shifted by even only one bin compared to the numerical BSD cutoff, the resulting difference between the two BSDs in this bin will dramatically increase the value of the mean squared error, possibly disfavoring a set of parameter values that might have well reproduced the numerical BSD for all the other bins. 
		
		To tackle both issues, we perform the fit using the mean squared error computed on max$\left( V^{2}\frac{dn_{b}}{dV},10^{-5}\right)$ in the interval $5\times 10^{-1}$ cMpc$^3\lesssim V \lesssim 10^{6}$ cMpc$^3$. We studied multiple other values of the $10^{-5}$ threshold and concluded that it does not change much the parameter inference result, only the absolute value of the error.\\
		
		In Fig.\ref{Fig:BSD_percolation}, we show the fit of the reference numerical BSD (gray histogram) at $\Xhii\sim 10\%$ (top) and $\Xhii\sim 22\%$ (bottom) using the logarithmic Softplus parameterization considering the dispersion with percolation (purple line) and without percolation (orange line). When using our percolation algorithm, the resulting BSD shape is closer to the numerical one and allows for a better fitting, especially of the sharp cut-off at high volumes. In Table \ref{Table:KhiBSD_percolation} we show the mean squared relative error of the BSD fitting process computed with the adaptations explained above, for the logarithmic Softplus parameterization using dispersion with and without using the percolation algorithm at various global ionization fraction. We can see that in both cases the mean squared error is always below 0.25 for $\Xhii > 10\%$ and that it is divided by a factor of $\sim 2$ when using percolation.\\

\begin{figure}[t]
\begin{center}
\includegraphics[width=\columnwidth]{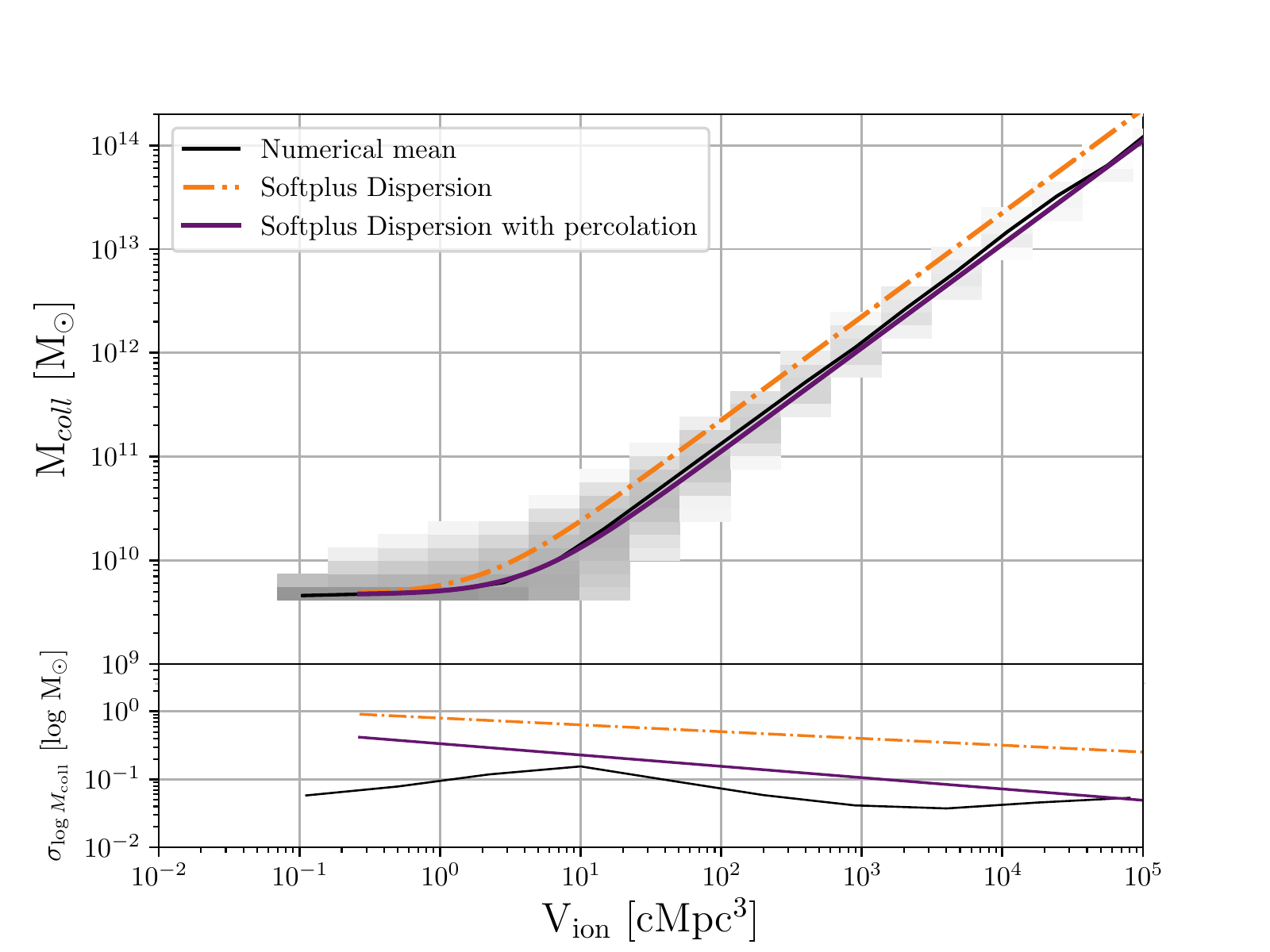}
\includegraphics[width=\columnwidth]{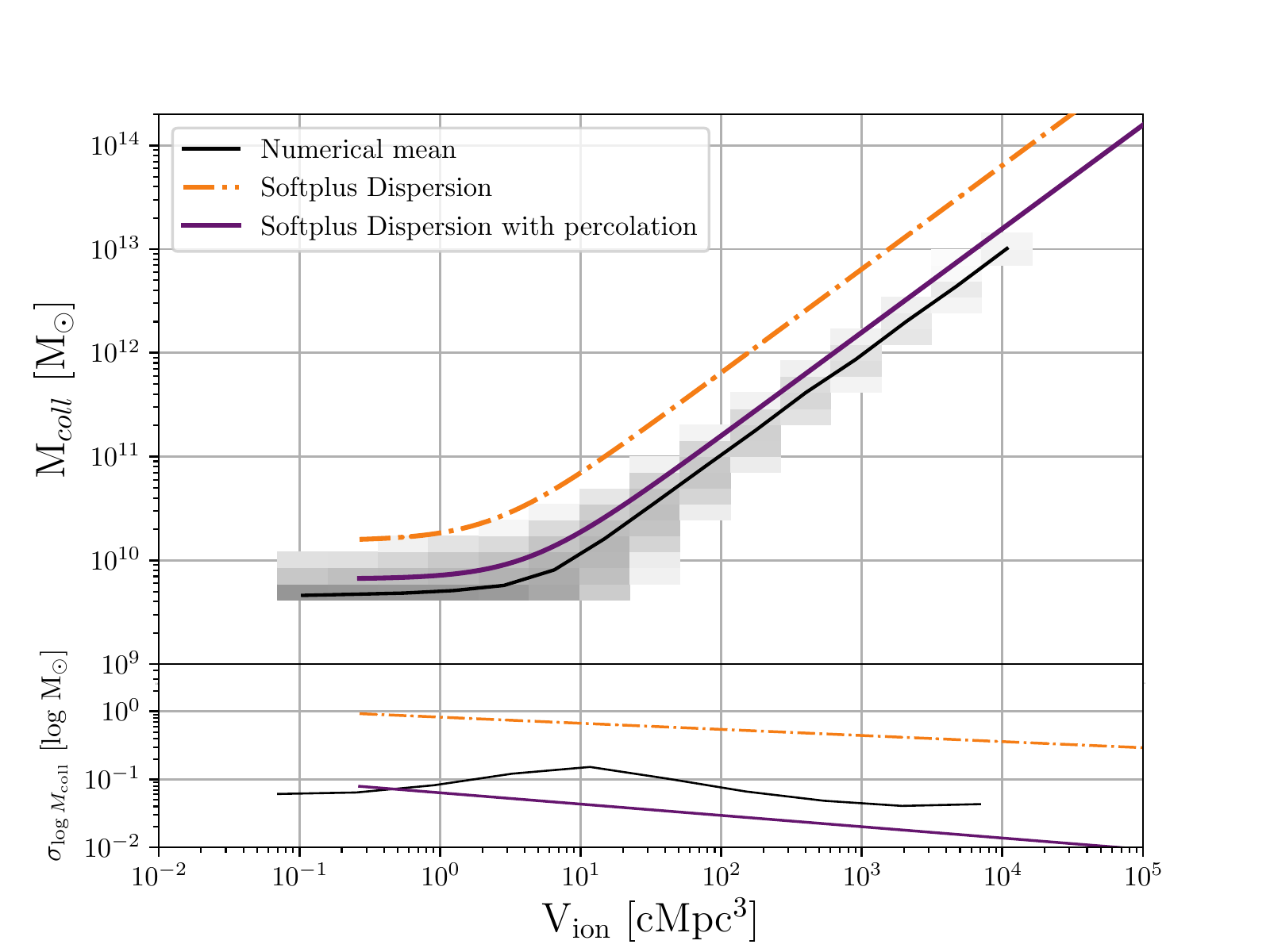}
\caption{Collapsed mass inside an ionized region as a function of its volume. The gray colormap represents the distribution of halos the HIRRAH-21 simulation and its numerical mean and dispersion are shown as black lines in the corresponding panels. The inferred relations using the logarithmic Softplus parameterization from Section \ref{sSec:SoftPlusRelation} considering the dispersion with percolation (purple line) and without percolation (orange line) are shown. The fit is done at $\Xhii\sim 10\%$ (top) and $\Xhii\sim 22\%$ (bottom).}
\label{Fig:M_of_V_percolation}
\end{center}
\end{figure}

\begin{table*}[t]
    \begin{center}
	\begin{tabular}{c | c | c | c | c | c | c}
	\hline
	\hline
	$\errMofV$ & $\Xhii\sim 10\%$ & $\Xhii\sim 16\%$ & $\Xhii\sim 22\%$ & $\Xhii\sim 30\%$ & $\Xhii\sim 38\%$ & $\Xhii\sim 50\%$ \\
    \hline
    Softplus Dispersion & 0.048 & 0.21 & 0.58 & 0.95 & 1.9 & 2.6 \\
    Softplus Dispersion + Percolation & 0.0030 & 0.003 & 0.040 & 0.14 & 0.29 & 0.63\\
	\hline
	\end{tabular}
	\caption{Mean squared error of the physical $M_{coll}(V_{\text{ion}})$ relation reconstruction for the logarithmic Softplus parameterization using dispersion with and without using the percolation algorithm at various global ionization fraction ($\Xhii\sim 10\%$, $16\%$, $22\%$, $30\%$, $38\%$, and $50\%$). }
\label{Table:KhiMofR_percolation}
\end{center}
\end{table*}

	This decrease in the mean squared error when fitting the BSD comes with a better parameter prediction ability compared to the case without percolation. In Fig.\ref{Fig:M_of_V_percolation} we show the collapsed mass inside an ionized region as a function of its volume at at $\Xhii\sim 10\%$ (top) and $\Xhii\sim 22\%$ (bottom). The gray colormap represents the distribution of halos the HIRRAH-21 simulation and its numerical mean and dispersion are shown as black lines in the corresponding panels. We also show the inferred relations using the logarithmic Softplus parameterization considering the dispersion with percolation (purple line) and without percolation (orange line). We see that both the inferred relation $M_{coll}(V_{\text{ion}})$ and its dispersion are significantly closer to the numerical ones when using percolation. To quantify this improvement, in Table \ref{Table:KhiMofR_percolation} we present the mean squared error of the physical $M_{coll}(V_{\text{ion}})$ relation reconstruction at various $\Xhii$. We see that the error on the inferred $M_{coll}(V_{ion})$ relation in the case with percolation is lower by an order of magnitude for $\Xhii\lesssim 30\%$ and by a factor $\sim 5$ for $30\%\lesssim\Xhii\lesssim 50\%$ compared to the error in the case without percolation.\\
	
	However, in Fig.\ref{Fig:M_of_V_percolation} and Table \ref{Table:KhiMofR_percolation} we also see that the mean squared error is increasing with increasing $\Xhii$.
	This is not surprising as, for example, for large average ionization fractions, the percolation increasingly involve the non-spherical percolated cluster, deviating from the simple sphere overlap considered in our model. Nonetheless, until $\Xhii\sim 30\%$ our percolation algorithm still allows to infer the physical $M_{coll}(V_{\text{ion}})$ relation with an mean squared error lower than $15\%$.

\begin{figure}[t]
\begin{center}
\includegraphics[width=\columnwidth]{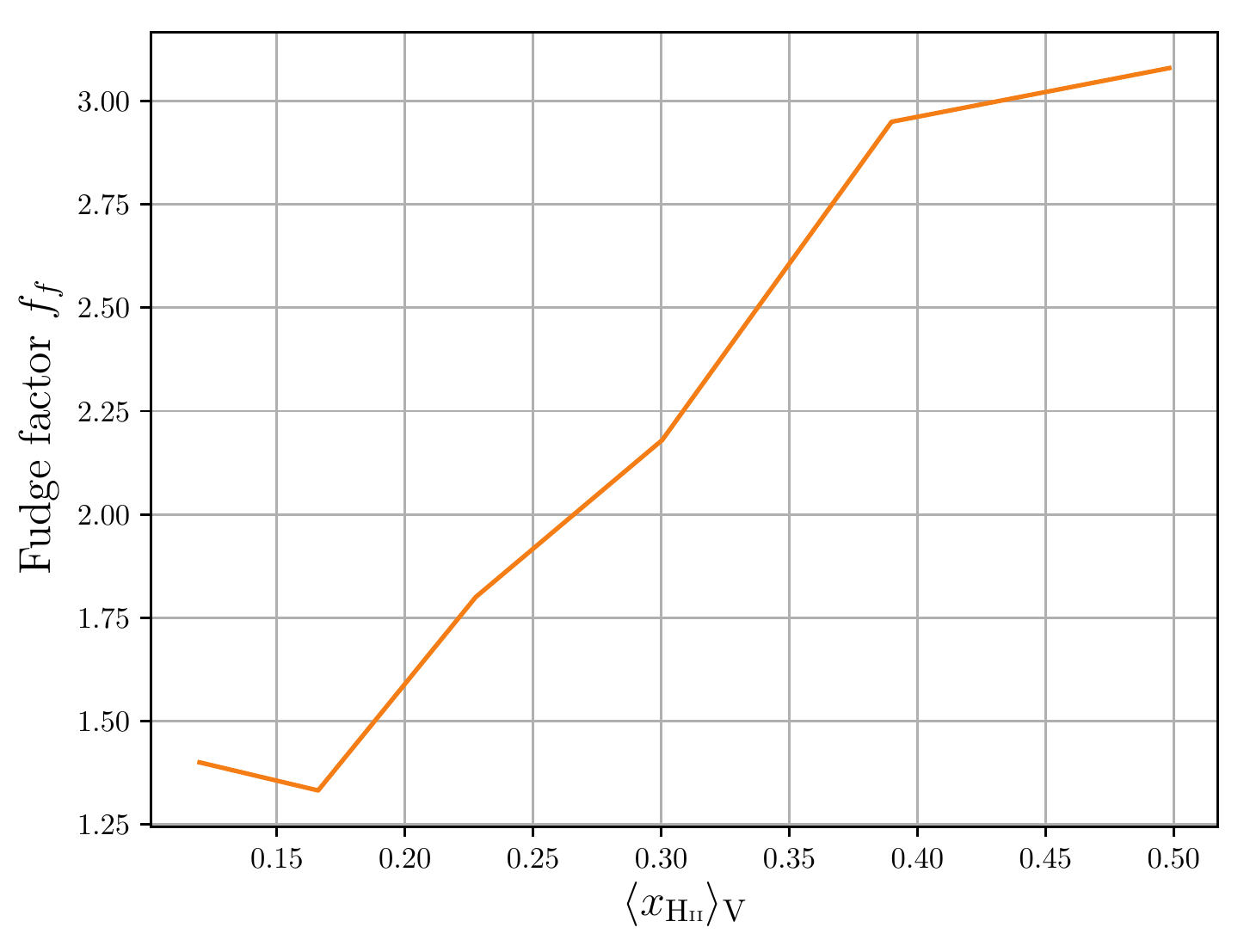}
\caption{Evolution of the fudge factor $f_{f}$ as a function of the volume-weighted global ionization fraction $\langle\Xhii\rangle_{\text{V}}$.}
\label{Fig:fudgeFactorEvolution}
\end{center}
\end{figure}

    In Fig.\ref{Fig:fudgeFactorEvolution} we represent the evolution of the fudge factor (resulting from the fitting process) as a function of the global ionization fraction $\Xhii$. It appears to increase more or less linearly on the studied range of $0.10\lesssim\Xhii\lesssim 0.50$. \red{One motivation for introducing this fudge factor is to correct for the fact that the positions of the ionized bubbles are in reality not uncorrelated, as is assumed in our simple percolation model. Thus it is not surprising that we need higher values in situations where bubble overlap is more present. Another contribution to the fudge factor may be the departure from the assumed spherical geometry, which is also getting stronger as the percolation process progresses.}

\section{Conclusions}\label{sec_ccl}

	In this work, we develop a new analytical model to compute the ionized bubble size distribution of the Epoch of Reionization ($5\lesssim z\lesssim 15$) starting from the underlying physical relation between the collapsed mass inside an ionized region and its size. We compare it to both, the fiducial model of \citet{Furlanetto2004a} and numerical results from the HIRRAH-21 simulation for global ionization fractions of $\Xhii\sim 1\%,3\%,10\%,16\%,22\%,30\%,38\%,$ and $50\%$. The HIRRAH-21 simulation is composed of 2048$^{3}$ particles in a 200 $h^{-1}$cMpc box and was performed using the \textsc{LICORICE} code, which fully couples the hydrodynamics with the radiative transfer of ionizing UV and X-rays. The key features and results of our models are :

\begin{enumerate}
	\item The underlying physical relation between the collapsed mass inside an ionized region and its volume is totally parameterizable, whereas in the original model by \citet{Furlanetto2004a} it takes a fixed form that enables an analytical computation. In our model, One can use any functional form and obtain the resulting BSD. In this study we use two functional forms, a power law and a logarithmic softplus, both producing a BSD in strong agreement with the numerical BSD directly computed from the HIRRAH-21 simulation before the percolation regime at $\Xhii\lesssim 10\%$.
	\item More than just the functional form, our model is largely adaptable to different theoretical frameworks. As it is based on the halo conditional mass function theory, it can use various mass function formalisms such as \citet{Press1974} and \citet{Sheth1999} formalisms\red{. It could even accommodate a numerically determined conditional mass function so that no biases are introduced}. One can choose to include the sample variance effect if needed. A functional form for the dispersion in the underlying $M_{coll}(V_{ion})$ physical relation can also be implemented.
	\item Using various algorithms, our model can infer the $M_{coll}(V_{ion})$ relation from the observed BSD. Before the percolation regime, for our two choices of parameterization, the resulting physical relation is in accordance with the relation computed from the HIRRAH-21 simulation, thus demonstrating the inference capability of the model. \red{Further studies using results from other simulation codes would be useful to better assess this inference capability.}
	\item We also develop a percolation algorithm that can apply a percolation effect to an existing BSD computed without percolation. \red{Good performances rely on introducing a fudge factor whose optimal value seems to be dependent on the average ionization fraction, although with a simple linear relation in the studied range.}  Using this algorithm, our model keeps a strong inference capability even in the percolation regime as it can recover the physical relation with a \red{mean squared} error lower than $1\%$ for $\Xhii\lesssim 16\%$ and lower than $15\%$ for $\Xhii\lesssim 30\%$ for the softplus parameterization. The inference capability decreases with increasing global ionization as the underlying approximations of our percolation algorithm likely do not stand at the latest stage of percolation.
\end{enumerate}

Let us emphasize again that the $M_{coll}(V_{ion})$ parameterization can be readily interpreted in term of underlying physical processes. For example in the logarithmic softplus parameterization, $M_{coll,th}$ is essentially the minimum halo mass for efficient star formation and, if $ \alpha \sim 1$, $V_{th}$ should be proportional to the amount of radiation emitted per unit collapsed mass. Thus constraining $M_{coll}(V_{ion})$ really provides information on the astrophysics. 

It is necessary to mention that the effect of thermal noise, imperfect calibration and foreground removal and limited angular resolution will impact the quality of the tomographic data from which an observed BSD will be derived. \red{The limited angular resolution of real observations will at least result in a cut-off of the BSD on small scales that will limit the range of scales that can be used in the inference process. It will also impact the general shape of the BSD on larger scales \citep{Giri2018}. It is possible that an addition could be made to our model to include the effect of angular resolution with a similar approach as for the effect of percolation but, at this stage, this remains an open question.
A variance in each bin of the observed BSD resulting from various sources, starting with the thermal noise, could be taken into account by using a $\chi^2$ evaluation for the goodness-of-fit.}

\red{A more difficult issue is the current domain of validity of the model (i.e. $\Xhii\lesssim 30\%$ at best). This likely corresponds to redshifts $z\gtrsim 9$ while the signal-to-noise ratio of observations with radio interferometers like SKA degrades with increasing redshift. One way to tackle this issue would be to devise a more sophisticated percolation model that would describe more precisely the spatial correlation in the ionised bubble distribution. Then our BSD model would likely be valid at higher ionization fractions and so at lower redshifts. In any case, }the inferred parameters will then have posterior distributions with potentially large variance. Quantifying this and comparing to the constraints that can be derived from other quantities such as the signal power spectrum will be necessary.

\begin{acknowledgements}
This work was performed using HPC resources from GENCI-CINES (grant 2018-A0050410557).
\end{acknowledgements}

%
%

\bibliographystyle{aa}
\bibliography{These,myref}
\end{document}